\documentclass[useAMS,usenatbib,letterpaper]{mn2e}
\usepackage{graphicx,amsmath,color,amssymb}
\usepackage[pdftitle={Reproducing the Kinematics of DLAs},pdfauthor={Simeon Bird}]{hyperref}
\usepackage[all]{hypcap}
\usepackage{subfigure}
\usepackage[utf8]{inputenc}

\newlength{\narrowFigurewidth}
\newlength{\Figurewidth}
\newlength{\wideFigurewidth}
\newcommand{\etal}
  {et al.}

\newcommand{\Lya}{Lyman-$\alpha\;$}
\newcommand{\Msun}{\, h^{-1} M_\odot}

\newcommand{\NHunit}{cm$^{-2}$}

\newcommand{\Mpch}{\, h^{-1} \mathrm{Mpc}}
\newcommand{\kpch}{\, h^{-1}\mathrm{kpc}}

\newcommand{\kms}{km~s$^{-1}$}

\newcommand{\cloudy}{{\small CLOUDY\,}}
\newcommand{\arepo}{{\small AREPO\,}}

\newcommand{\NHI}{N_\mathrm{HI}}
\newcommand{\velw}{$\Delta \mathrm{v}_{90}$}
\newcommand{\velwm}{\Delta \mathrm{v}_{90}}
\newcommand{\vvir}{\mathrm{v}_\mathrm{vir}}
\newcommand{\fedg}{f_\mathrm{edg}}
\newcommand{\fmm}{f_\mathrm{mm}}

\newcommand{\SiII}{Si$\mathrm{II}$}

\title[Reproducing the Kinematics of DLAs]{Reproducing the Kinematics of Damped Lyman-$\alpha$ Systems}
\author[S. Bird \etal]{Simeon Bird$^{1}$\thanks{E-mail: sbird@andrew.cmu.edu}, Martin Haehnelt$^{2}$, Marcel Neeleman$^{3}$, Shy Genel$^{4}$, 
\newauthor Mark Vogelsberger$^{5}$, Lars Hernquist$^{4}$ 
\vspace{4mm}\\
$^1$Institute for Advanced Study, 1 Einstein Drive, Princeton, NJ, 08540, USA\\
$^2$Institute for Astronomy, Cambridge, CB3 0HA, UK \\
$^3$Department of Physics and Center for Astrophysics and Space Sciences, UCSD, La Jolla, CA 92093, USA \\
$^4$Harvard-Smithsonian Center for Astrophysics, 60 Garden Street, Cambridge, MA 02138, USA \\
$^5$MKI and Department of Physics, Massachusetts Institute of Technology, 77 Massachusetts Avenue, Cambridge, MA 02139, USA \\
}

\begin{document}

\maketitle

\begin{abstract}
We examine the kinematic structure of Damped \Lya Systems (DLAs) in a series of cosmological hydrodynamic simulations using the \arepo~code.
We are able to match the distribution of velocity widths of associated low ionization metal absorbers substantially better than earlier work. Our simulations produce a 
population of DLAs dominated by haloes with virial velocities around $70$ \kms, consistent with a picture of relatively small, 
faint objects. In addition, we reproduce the observed correlation between velocity width and metallicity and the equivalent width distribution of \SiII. 
Some discrepancies of moderate statistical significance remain; too many of our spectra show absorption concentrated at the edge of the profile
and there are slight differences in the exact shape of the velocity width distribution.
We show that the improvement over previous work is mostly due to our strong feedback from star formation and our detailed modelling of the metal ionization state.
\end{abstract}


\section{Introduction}

Columns of neutral hydrogen with $\NHI > 10^{20.3}$ \NHunit~are referred to as Damped \Lya Systems (DLAs), 
heavily saturated \Lya~absorption features in quasar spectra \citep{Wolfe:1986}.
At these densities, hydrogen is self-shielded from the ionizing effect of the diffuse radiation
background and so is neutral \citep{Katz:1996a}.
Since the \Lya line is saturated, kinematic information on the absorbing gas is 
extracted from low ionization metal lines such as \SiII, 
under the assumption that these metals trace the cold neutral gas.
Particularly important is the low ionization velocity width, customarily defined as the length of 
spectrum covering $90\%$ of the total optical depth across the absorber \citep{Prochaska:1997}.
The velocity width provides indirect information on the virial velocity of the DLA host halo.
The distribution of velocity widths exhibits a peak at $70$\kms~\citep{Prochaska:1997, Neeleman:2013}, 
and a long tail to higher velocities, which, together with the large number of edge-leading spectra, 
has been interpreted using simple semi-analytic models to suggest that
DLAs occur predominantly in large rotating discs \citep{Prochaska:1997, Jedamzik:1998, Maller:2001}
\footnote{Although the kinematics and chemistry of metal-poor DLAs are suggestive of Milky Way dwarf spheroidals \citep{Cooke:2014}.}.

By contrast, the results of cosmological simulations suggested that DLAs are primarily hosted in 
small protogalactic clumps \citep{Haehnelt:1998, Okoshi:2005}, and large velocity widths can arise from systems aligned
along the line of sight \citep{Haehnelt:1996}.
However, most previous simulations have been unable to match the observed velocity width distribution, producing a distribution 
which peaks at too low a value. It has been suggested that this discrepancy can be alleviated if 
strong stellar feedback induces a high virial velocity cutoff below which DLAs cannot form \citep{Barnes:2009, Barnes:2014}.
The recent publication of an updated and expanded catalogue of velocity widths \citep[henceforth N13]{Neeleman:2013}, and the detection of a correlation
between velocity width and metallicity \citep{Ledoux:2006, Prochaska:2008, Moller:2013, Neeleman:2013} make it timely to re-examine this question.

In this paper, we compare cosmological hydrodynamic simulations first presented in \cite{Bird:2014} 
to the observations of N13. In \cite{Bird:2014} we examined the column density function, DLA metallicity, and DLA bias, 
finding generally good agreement between observations and a model with strong stellar feedback. Our model 
produced DLAs hosted in haloes with a relatively modest mean mass of $10^{10} - 10^{11}\Msun$.
One remaining problem with the model is that it produced an excess of HI at $z < 3$. 
We do not expect this discrepancy to significantly affect our results as the HI column density is uncorrelated
with the velocity width \citep{Neeleman:2013} and our results are quoted at $z=3$, the mean redshift 
of the observed velocity width sample.
Here we examine the distribution of velocity widths, the correlation between velocity width and metallicity, the spectral edge-leading statistics and 
the equivalent width distribution of the \SiII~$1526$~\AA~line. We show that some of our models achieve good 
agreement with the observations, and explain what features of our modelling are important in producing this effect.

\cite{Barnes:2014} noted a difference between the velocity width distribution as measured by \cite{Wolfe:2005} and N13,
despite both distributions being drawn from a similar sample. This occurs for two reasons.
First, care was taken in the N13 sample to remove DLAs selected using previously known
metal lines, as these systems would have a higher metallicity and thus velocity width. 
Secondly, N13 used only very high resolution spectra (FWHM $\sim 8$ \kms), as lower resolution spectra 
cannot resolve the smallest velocity width systems, and again bias the velocity width distribution high. 

Several previous simulations have been used to examine the velocity width distribution, beginning with those of 
\cite{Haehnelt:1998}, who were able to reproduce the shape of the absorption profiles using hydrodynamic simulations 
of four galactic haloes. By including a larger sample of haloes and strong supernova feedback, \cite{Pontzen:2008}
were able to reproduce the column density function, DLA metallicity distribution, and, in \cite{Pontzen:2010}, the 
properties of analogous objects in the spectra of gamma ray bursts. \cite{Razoumov:2009} examined the kinematic distribution of DLA metal lines in
AMR simulations of three isolated haloes. Using adaptive mesh refinement simulations of two regions in extreme 
density environments, \cite{Cen:2012} was able to bracket the observed distribution of velocity widths.
Finally, fully cosmological simulations were performed by \cite{Tescari:2009}, who also examined the velocity width distribution 
for a variety of supernova feedback models. 

We use cosmological simulations. DLAs probe random locations 
lit up by quasars and simulations of DLAs should thus select sight-lines at random positions in a cosmological volume.
Obtaining representative samples from simulations of isolated haloes is therefore not straightforward and may produce biased or misleading results.
Our models for subgrid supernova and AGN feedback are based on the framework 
discussed in \cite{Vogelsberger:2013, Torrey:2013}, and applied in \cite{Vogelsberger:2014b, Vogelsberger:2014} and \cite{Genel:2014}.
We use the moving mesh code \arepo\ \citep{Springel:2010} to solve the equations of hydrodynamics.
DLAs arise from self-shielded gas, and so radiative transfer effects must be taken into account.
We use the prescription of \cite{Rahmati:2013a} to model radiative transfer effects in neutral gas, but we should note that they have also been examined by
e.g.~\cite{Pontzen:2008, FaucherGiguere:2010, Fumagalli:2011, Altay:2011, Yajima:2011} and \cite{Cen:2012}.

\section{Methods}

Before explaining our analysis of the artificial spectra, we shall briefly summarize our simulations, 
described in detail in \cite{Bird:2014}. Readers familiar with the simulations may wish to skip to Section \ref{sec:ionfrac}.

We use the moving mesh code \arepo\ \citep{Springel:2010},
which combines the TreePM method for gravitational interactions with a moving mesh hydrodynamic solver.
Each grid cell on the moving mesh is sized to contain a roughly fixed amount of mass, and to move with 
the bulk motion of the fluid; gas and metals advect between grid cells to account for local discrepancies.
We assume ionization equilibrium, and account for optically thick gas as described in Section~\ref{sec:hishield}.
We include the heating effect of the UV background (UVB) radiation following the model of \citet{Faucher:2009}.
In some of our simulations the UVB amplitude has been increased by a factor of two to agree with the latest intergalactic medium (IGM)
temperature measurement of \cite{Becker:2010}, although this does not affect our results. 
Star formation is implemented with the effective two-phase model of \cite{Springel:2003}.

Our initial conditions were generated at $z=127$ from a linear theory power spectrum 
using cosmological parameters consistent with the latest WMAP results \citep{WMAP}.
The box size is $25 \Mpch$, except for one simulation used to 
check convergence. Each simulation has $512^3$ dark matter (DM) particles, and 
$512^3$ gas elements, which are refined and de-refined as necessary to 
keep the mass of each cell roughly constant in the presence of star formation. 

\subsection{Feedback Models}
\label{sec:feedback}

Feedback from star formation is included following the model described in detail in \cite{Vogelsberger:2013}.
Star-forming ISM cells return energy to the surrounding gas by stochastically creating wind particles with an energy of 
$\mathrm{egy}_\mathrm{w} = 1.09\; \mathrm{egy}_\mathrm{w}^0$, where
$\mathrm{egy}_\mathrm{w}^0 = 1.73 \times 10^{49} \mathrm{erg} \,M_\odot^{-1}$ is the expected available
supernova energy per stellar mass. Wind particles, once created, interact gravitationally 
but are decoupled from the hydrodynamics until they reach a density threshold or a maximum travel time. 
They are then dissolved and their energy, momentum, mass and metal contents are added to the gas cell at their current locations. 
We assume that the mass loading of the winds, $\eta_\mathrm{w}$, is given by 
$\eta_\mathrm{w} = 2 \mathrm{egy}_\mathrm{w} / v_\mathrm{w}^2$, where $v_\mathrm{w}$ is the 
wind velocity. $v_\mathrm{w}$ is chosen to scale with the local DM velocity 
dispersion, $\sigma^\mathrm{1D}_\mathrm{DM}$, which correlates with the maximum DM circular 
velocity of the host halo \citep{Oppenheimer:2008}. Thus we define
\begin{equation}
 v_\mathrm{w} = \kappa_\mathrm{w} \sigma^\mathrm{1D}_\mathrm{DM}\,.
\end{equation}
Our wind model produces very large mass loadings in small haloes, which 
allows it to roughly match the galaxy stellar mass function at $z=0$ \citep{Okamoto:2010, Puchwein:2013}.

\begin{table}
\begin{center}
\begin{tabular}{|l|c|c|c|l|}
\hline
Name & $v^m_w$ & AGN & $\kappa_\mathrm{w}$ & Notes \\
\hline 
HVEL     &  600     & Yes &  $3.7$ & \\ 
DEF     &  0       & Yes &  $3.7$ & $2\times$ UVB amplitude \\ 
SMALL     &  200       & Yes &  $3.7$ & $10 \Mpch$ box \\ 
FAST     &  0       & Yes &  $5.5$ & \\ 
NOSN     &  -       & No  &  -     & No feedback \\ 
\hline
\end{tabular}
\end{center} 
\caption{Table of simulation parameters. $v^m_w$ gives the minimum wind velocity in \kms, and $\kappa_\mathrm{w}$ the specific wind energy.
We omit some simulations from \protect\cite{Bird:2014} which give results similar to those shown, as explained in the text.
Note that in \protect\cite{Bird:2014} the DEF simulation is called 2xUV.}
\label{tab:simulations}
\end{table}

The model parameters we consider are shown in detail in Table \ref{tab:simulations}.
DEF uses the reference model of \cite{Vogelsberger:2013}, with the amplitude of 
the UVB doubled to match the results of \cite{Becker:2013}. HVEL adds a minimum wind velocity, which has the effect of significantly suppressing
the abundance of DLAs in haloes smaller than $10^{11} \Msun$. NOSN shows the results without supernova feedback, and FAST shows the effect of increasing
the wind velocity by $50\%$, while decreasing the mass loading to keep the wind energy constant. This reduces the star 
formation rate, but allows the simulation to better match $\Omega_\mathrm{DLA}$ at $z < 3$.
While we include AGN feedback \citep{DiMatteo:2005, Springel:2005f}, it does not significantly affect our results, as our 
DLA population is dominated by relatively low-mass haloes.
The other simulations mentioned in \cite{Bird:2014}, and omitted here, yield results very similar to DEF.
We used a $10 \Mpch$ box to check the convergence of our results with respect to resolution. This simulation had a minimum wind velocity of $200$ \kms, 
but we checked this did not affect our results by comparing the results from a $25 \Mpch$ box with this parameter to DEF.

\subsection{Metals}
\label{sec:metals}

Star particles in our simulation produce metals which enrich the surrounding gas. Enrichment events are assumed to occur from asymptotic giant branch stars
and supernovae, with the formation rate of each calculated 
using a \cite{Chabrier:2003} initial mass function. The elemental yields of each type of mass return are 
detailed in \cite{Vogelsberger:2013}. Metals are distributed into the gas cells 
surrounding the star using a top-hat kernel with a radius chosen to enclose a total mass equal to $256$ times 
the average mass of a gas element. We have checked that distributing the metal within a radius enclosing $16$ times the 
gas element mass does not affect our results. \arepo~is formulated to allow metal species to advect self-consistently 
between gas cells, naturally creating a smooth metallicity distribution and allowing the gas to diffuse through the DLA. 
Furthermore each DLA is enriched by multiple stars, and the enrichment radii of individual star particles frequently overlap. 
The metallicity of each wind particle is a free parameter, given by
$Z_\mathrm{w} = \gamma_\mathrm{w} Z_\mathrm{ISM}$. We adopt $\gamma_\mathrm{w} = 0.4$, following \cite{Vogelsberger:2013},
but we have checked that this free parameter does not affect the distribution of metals within DLAs \citep{Bird:2014}, whose 
metallicity is dominated by local enrichment.

\subsection{Hydrogen Self-Shielding}
\label{sec:hishield}

The high density gas that makes up DLAs self-shields from the ionizing effects of the UVB.
We model the onset of self-shielding using the fitting function from  \cite{Rahmati:2013a}. 
The photo-ionization rate is assumed to be a function of the gas temperature and the hydrogen density, shown in Figure \ref{fig:ionfrac}.
A surprisingly good approximate model is to assume a sharp transition to neutral gas at $n_\mathrm{H} \sim 10^{-2}$ cm$^{-3}$ \citep{Haehnelt:1998}.
Star forming gas is assumed to have a temperature of $10^4$ K, the temperature of the gas at the star formation threshold density,
and is thus almost completely neutral.
We include the self-shielding model into the simulation itself, rather than applying it in post-processing. 
This allows the cooling rate of high-density gas to be computed under the assumption that it is neutral, making 
the gas temperature self-consistent.
We checked that the effect this has on the velocity width distribution is negligible,
although it does reduce the mean \SiII~equivalent width by about $0.2$ dex.
We neglect the effect of local stellar radiation and account for molecular hydrogen formation as described in \cite{Bird:2014}, although
as we use the HI column density only to identify DLAs this does not affect our results.

\subsection{Metal Ionisation}
\label{sec:ionfrac}

We use \cloudy~version 13.02 \citep{Ferland:2013} to compute the ionization fraction of metal species in the presence of an 
external radiation background.
\cloudy~is run in single-zone mode, which assumes that the density and temperature are constant within each gas cloud.
We thus neglect density and thermal structure on scales smaller than the resolution element of each simulation, which is consistent 
as this structure would by definition be unresolved. 

Self-shielding at high densities is included by multiplying the amplitude of the external background radiation by an energy-dependent effective self-shielding factor $S(E)$.
For $E < 1$ Rydberg (the Lyman limit), $S(E) = 1$, corresponding to no shielding.
For higher energies, we compute the self-shielding factor $S(E)$ from the 
self-shielded hydrogen photo-ionization rate, $\Gamma_\mathrm{SS} (n, T)$, by
\begin{equation}
S(E) = \frac{\Gamma_\mathrm{SS} (n'(E), T)}{\Gamma_\mathrm{UVB}}\,.
\end{equation}
Here, $T$ is the gas temperature, and $n'$ is an effective density to ionizing photons, used to correct for the dependence
of hydrogen cross-section on energy. If $\sigma_\mathrm{HI} (E)$ is the hydrogen cross-section \citep{Verner:1996}, 
the optical depth from neutral hydrogen is
\begin{equation}
\tau \sim n\, \sigma_\mathrm{HI}( E )\,,
\end{equation}
giving an effective density of 
\begin{equation}
n'(E) = \frac{n\, \sigma_\mathrm{HI} (E)}{\sigma_\mathrm{HI} (1 \mathrm{Ryd})}\,.
\end{equation}
Note that for $E \gg 1$, $\sigma_\mathrm{HI} (E) \sim E^{-3}$. Thus
at high energies the effect of self-shielding is reduced.\footnote{The python module for generating our cloudy tables is publicly available at \url{https://github.com/sbird/cloudy_tables}.}

\begin{figure}
\includegraphics[width=0.45\textwidth]{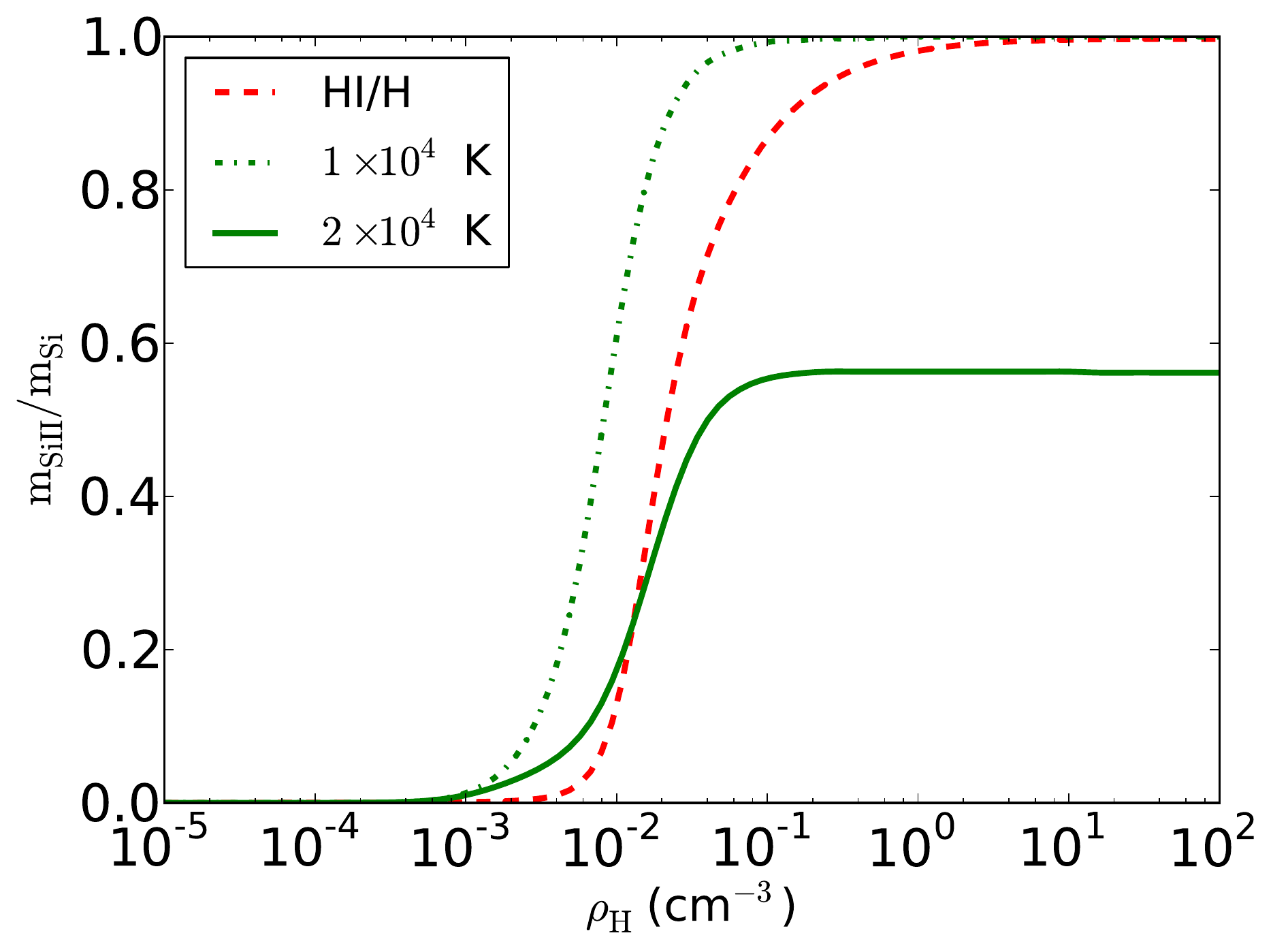}
\caption{The mass fraction of silicon in \SiII~as a function of density at $z=3$, for
$10^4$K (dot-dashed green) and $2\times 10^4$K (solid green).
The red dashed line shows the neutral fraction of hydrogen at $10^4$K, for comparison.}
\label{fig:ionfrac}
\end{figure}

Figure \ref{fig:ionfrac} shows the fraction of silicon in \SiII~as a function of density. At a temperature of $10^4$K, 
the transition to \SiII~from higher ionization states takes place at similar densities to the transition to neutral hydrogen.
However, because the ionization potential of \SiII~($16.3$ eV) is slightly higher than that for HI ($13.6$ eV),
\SiII~persists in slightly lower density gas.
We computed the HI-weighted temperature for each DLA spectrum, and found that it was almost always between $10000$ and $12000$ K, 
independently of column density.

We have checked the effect of using different models for self-shielding. 
Neglecting the dependence of the cross-section on energy for $E > 1$ Ryd, so that
\begin{equation}
S(E) = \frac{\Gamma_\mathrm{SS} (n, T)}{\Gamma_\mathrm{UVB}}\,,
\end{equation}
had negligible effect, as the ionization potential for \SiII~is relatively close to the Lyman limit.
Neglecting self-shielding altogether, $S(E) = 1$, had similarly little effect on the velocity width, 
but reduced the mean equivalent width. The velocity width is by construction 
insensitive to the total ionic abundance. Instead it is sensitive to places where the ionic abundance sharply decreases. 
If self-shielding is (unphysically) neglected, changes in \SiII~ionic fraction occur in regions where the temperature 
or metallicity of the gas changes. Neglecting self-shielding entirely does not affect our velocity width results because for our self-shielding 
prescription, the edge of the self-shielded gas also corresponds to an increase in gas temperature and a decrease in gas metallicity.
A self-shielding prescription with a higher density threshold, as in \cite{Tescari:2009}, will create an inner bubble of high \SiII~fraction 
inside the cool dense gas, which will dominate the absorption and reduce the velocity width.
Finally, we tested the assumption, following \cite{Pontzen:2008}, that $n(\mathrm{SiII})/n(\mathrm{Si}) = n(\mathrm{HI})/ n(\mathrm{H})$, which, 
as described in Appendix \ref{ap:compare}, creates a moderate bias towards smaller velocity widths.

When generating simulated spectra, the ionization fraction is computed for each gas cell using a lookup table. 
We have checked that our lookup table is populated sufficiently finely that the error from interpolation is negligible.

\subsection{Sightline Positions}
\label{sec:sightline}

The position of quasars on the sky should be uncorrelated with foreground absorption. 
Thus, to make a fair comparison to observations, sight-lines must be chosen to produce a fair volume-weighted sample 
of the total DLA cross-section. We achieve this by placing sight-lines at random positions in our box, computing the total HI 
column density along the sight-line, and selecting those with total column density above the DLA threshold, $\NHI > 10^{20.3}$ \NHunit.
We have checked that our results are not affected if we instead use the total column density within $\pm 500$ km/s
of the strongest absorption. As an optimization, we used the projected column density grid computed in \cite{Bird:2014}, and selected sight-lines 
only within grid cells which contain at least a Lyman-limit system (LLS; $\NHI > 10^{17}$ \NHunit). Using only those cells which contain a DLA
induces a slight bias, as DLAs are not perfectly aligned with grid cell boundaries.

\subsection{Artificial Spectra}
\label{sec:spectra}

Here we shall describe in detail our mechanism for generating the artificial spectra we use to compute our kinematic 
statistics.\footnote{Our implementation is available at \url{https://github.com/sbird/fake_spectra} and \url{https://github.com/sbird/vw_spectra}.}

We first compute for each gas cell a list of nearby sight-lines, using a sorted index table to speed up computation.
Particles which are not near any line are discarded. The ionization fraction of the remaining cells is computed as described in Section \ref{sec:ionfrac}.
The ionic density in each cell is interpolated on to the sight-line at a redshift space position
\begin{equation}
v = H(z) x + v_\mathrm{par}\,.
\label{eq:vpar}
\end{equation}
Here, $H(z)$ is the Hubble parameter, and converts from physical Mpc to~\kms. $x$ is the position of the cell in the box, and $v_\mathrm{par}$ the velocity
of the cell parallel to the line of sight. Thus we include redshift due to the cell motion.

The absorption along each sight-line is the sum of the absorption arising from each cell, computed by 
convolving the Voigt profile, $\mathcal{V}(v)$, with a density kernel. Thus
\begin{equation}
\tau = \mathcal{V}(v) \ast n\left(\sqrt{v^2/H^2(z)+r^2_\mathrm{perp}}\right)\,,
\end{equation}
where $v$ was defined in equation (\ref{eq:vpar}) and $n\left(\sqrt{v^2/H^2(z)+r^2_\mathrm{perp}}\right)$ is the smoothing kernel as a function of the distance 
from the cell centre to a point in the bin. These convolutions are computed by a simple numerical integration using 
the trapezium rule, and we neglect tails where $\tau < 10^{-5}$.
The smoothing length is chosen to be the radius of a sphere with the volume of the grid cell.
We have also considered using a top-hat kernel, and found that it changes the resulting 
$\tau$ by less than $10$\% and does not affect our results.

Our simulated spectra have a spectral resolution of $1$~\kms. We verified explicitly that our results 
were converged with respect to the spectral resolution. To better mimic the broadening function of the 
observational data, we convolved the simulation output with a Gaussian of FWHM $8$ \kms. 
Some of our DLA spectra (about $5\%$) had an \SiII~column 
density less than $2\times 10^{11}$ \NHunit, which would result in undetectable metal absorption. 
These spectra largely disappear in the higher resolution simulation, 
where only $0.1\%$ of sight-lines are metal poor, implying that they correspond to unresolved sites of star formation.
We checked that their inclusion does not affect our results.

\subsection{Kinematic Statistics}
\label{sec:kinematics}

We will examine three kinematic statistics extracted from DLA-selected \SiII~absorbers and originally defined in \cite{Prochaska:1997}. 
These are the velocity width, $v_{90}$, the mean median statistic, 
$\fmm$ and the edge leading statistic, $\fedg$. 
The velocity width includes $90\%$ of the integrated optical depth across the absorber
and is defined as 
\begin{equation}
 \velwm = v_\mathrm{high} - v_\mathrm{low}\,,
\end{equation}
where, if $F_\tau(v)$ is the fraction of the 
total optical depth in the absorber at $v$, 
$v_\mathrm{low}$ and $v_\mathrm{high}$ are chosen so that $F_\tau(v_\mathrm{low}) = 0.05$
and $F_\tau(v_\mathrm{low}) = 0.95$.

The mean-median statistic, $\fmm$, is the difference in units of the velocity width between the median velocity, $v_\mathrm{median}$, 
and the mean velocity, $v_\mathrm{mean}$. These are defined respectively by $F_\tau (v_\mathrm{median}) = 0.5$, and 
$v_\mathrm{mean} = v_\mathrm{low}/2 + v_\mathrm{high}/2$. Thus
\begin{equation}
 \fmm = \frac{|v_\mathrm{mean} - v_\mathrm{median} |} {\velwm / 2}\,.
\end{equation}
Lastly, the edge leading statistic, $\fedg$, is the difference in units of the velocity width between 
$v_\mathrm{peak}$, the velocity corresponding to the largest optical depth along the absorber, and the mean velocity.
Thus
\begin{equation}
 \fedg = \frac{|v_\mathrm{mean} - v_\mathrm{peak} |} {\velwm/2}\,.
\end{equation}

Observationally, these statistics are extracted from the strongest
unsaturated line, usually a transition of silicon or sulphur. To model this, we compute spectra for every transition of \SiII, and 
select the strongest transition with a peak optical depth between $3$ and $0.1$. These limits were chosen to roughly 
match the maximal $\tau$ in the observed sample.
In approximately $10$\% of our spectra, no \SiII~line lies within this band. In the observational sample a different ionic transition would be used, but 
since almost all simulated spectra have $\tau < 10$, we select the weakest transition with a peak optical depth greater than $0.1$.
Although in principle the different line broadening in each transition could affect the derived kinematics, we have checked that this 
does not occur in practice; our results are unchanged if we simply compute the velocity width from, e.g. \SiII~$1808$ \AA.

Since extended metal absorbers are unsaturated and often contain considerable structure, there is potential
ambiguity over the width of the velocity interval to consider a single absorber.
We follow N13 as closely as possible.
First, the location of the peak \SiII~absorption is considered to be the central point of the absorber; observationally this,
rather than the saturated \Lya~line, is used to determine the DLA redshift. We then consider the absorber to extend at least 
$500$ \kms~to either side of the peak. If there is still significant absorption in the (saturated) \SiII~$1260$ \AA~line
at the edges of the absorber, it is extended further. Within this search range, the absorber
is shrunk to the radius exhibiting significant optical depth in \SiII~$1260$ \AA.
Significant absorption is defined to be $4 \sigma$ in the presence of noise, or
$\tau > 0.2$ in noiseless spectra, anywhere in a $20$ \kms~bin. 
To check the impact of the observational choice of absorber size, we chose a minimum size of $\pm 250$ \kms. 
With this choice, the number of spectra with $\velwm \gtrsim 300$ \kms decreased by $\sim 1/3$, but spectra with smaller
velocity widths were not affected.
We emphasize that we have used the same definitions as in the analysis of the observed spectra. 
As final confirmation of our velocity widths, we performed a blind analysis; $100$ simulated spectra were analysed
with the observational pipeline, and the velocity width measurements from the two pipelines found to be consistent for every spectrum.

\subsection{Error Analysis}
\label{sec:errors}

The likelihood of our models giving rise to the observed data is assessed with bootstrap errors.
We choose $10000$ samples at random each containing $100$ simulated spectra, the same size 
as the N13 data set. The desired statistic and its distribution is computed, and binned in the same way as the data for each sample. 
We show as $1-\sigma$ errors the region including $68\%$ of the derived distributions.

Systematic errors are incorporated by adding a normally distributed random number to each derived statistic.
To calibrate the expected level of systematic error, $100$ simulated spectra were generated and analysed by one of us using the same pipeline 
as the real, observed, spectra, assuming a signal to noise ratio (SNR) of $10$. The difference between these velocity widths and the true velocity width derived
from the noiseless spectra gives an estimate of the systematic error due to noise.
In the case of the velocity width distribution, the systematic error is well-fitted by a Gaussian with standard deviation $5$ \kms and zero mean.
For $\fedg$ and $\fmm$, systematic error is negligible compared to variance induced by the limited sample and is thus neglected.

\section{Results}

\subsection{Velocity Widths}
\label{sec:velwidth}

\subsubsection{Matching the Observed Velocity Width Distribution}
\label{sec:velwidthmatch}

\begin{figure*}
\includegraphics[width=0.45\textwidth]{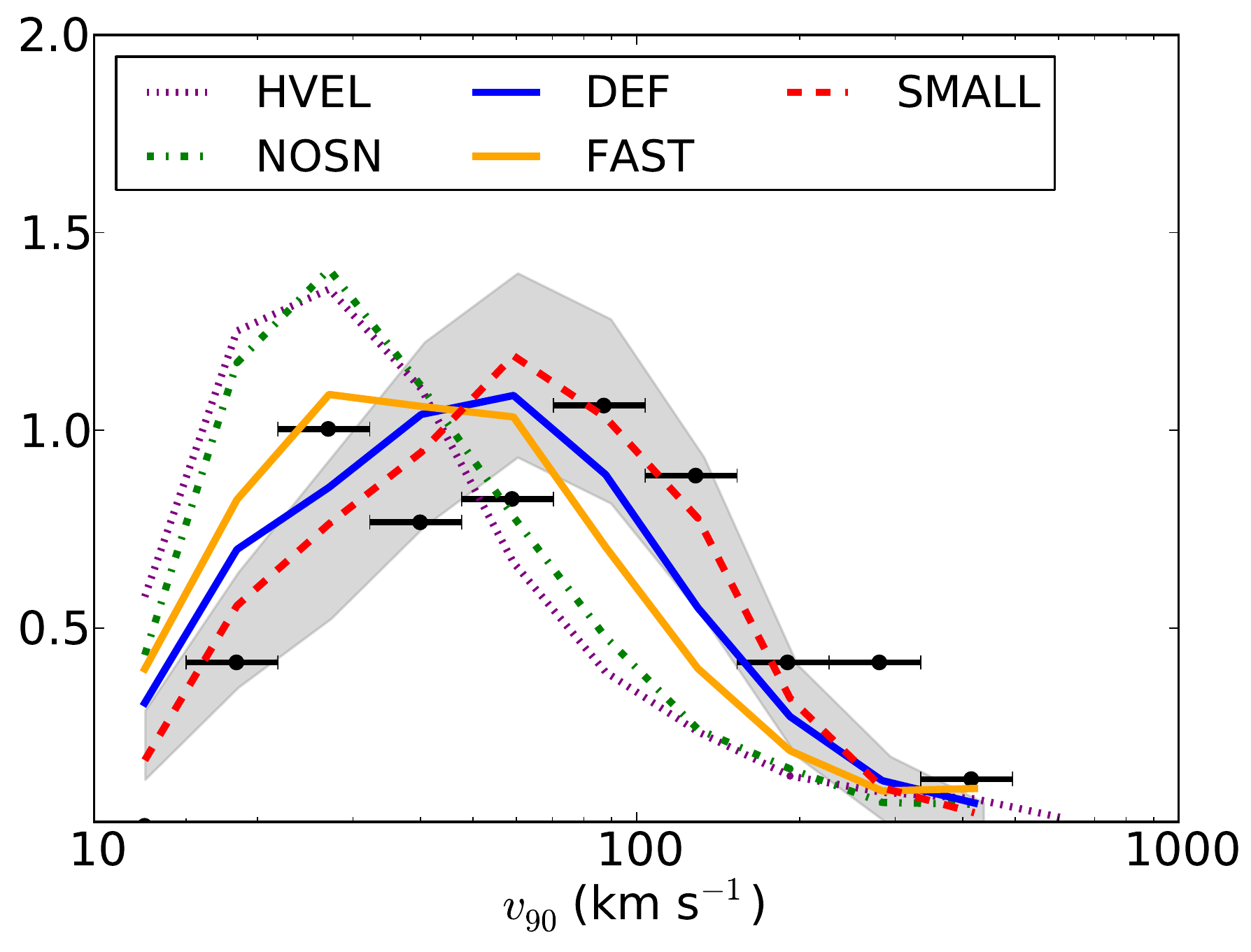}
\includegraphics[width=0.45\textwidth]{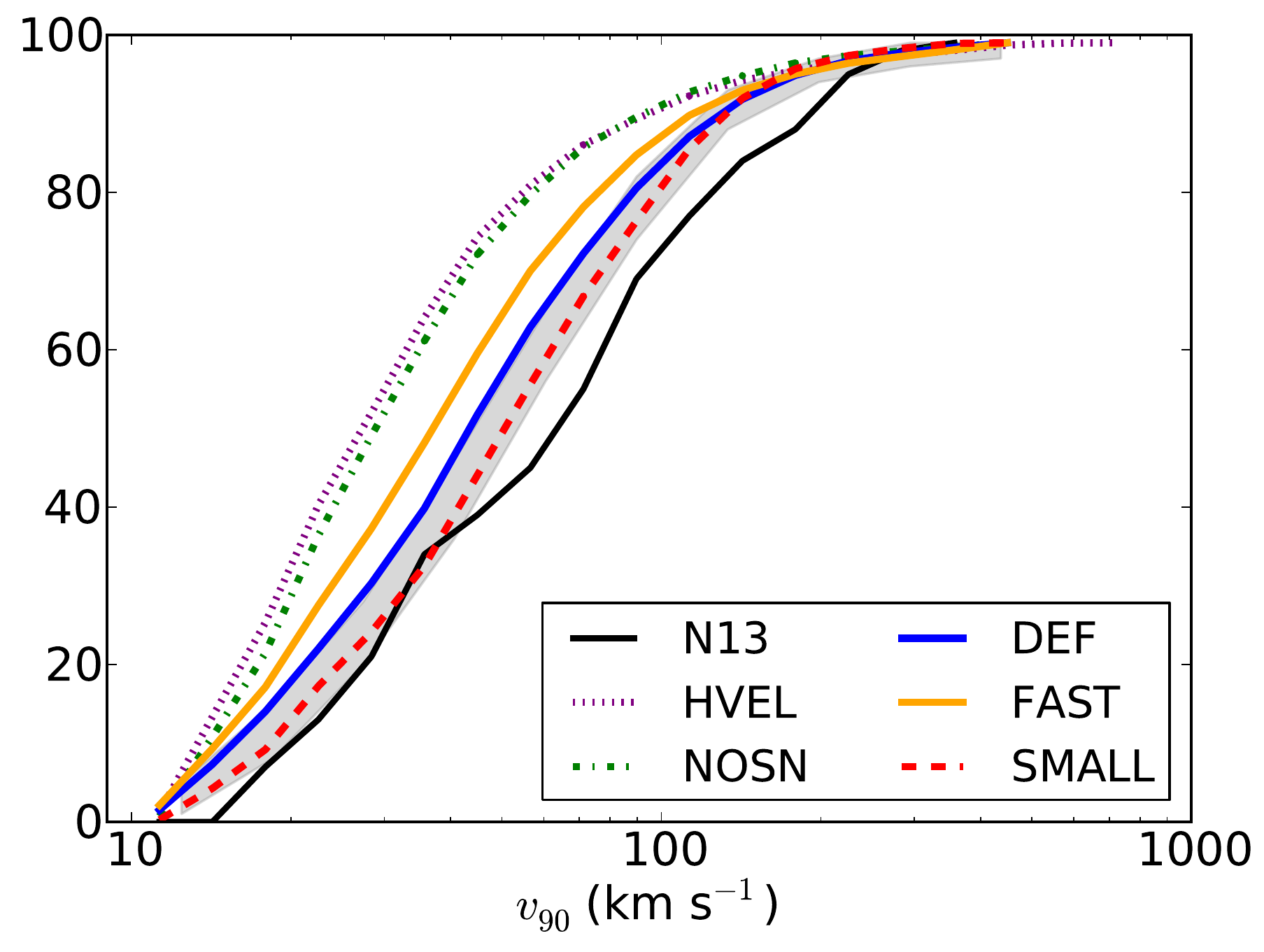}
\caption{Left: the distribution of velocity widths from our simulations at $z=3$, compared to the observed sample from N13 
over the combined redshift range. Right: the cumulative distribution function normalized to the size of the N13 sample, from the same simulations. 
Grey shaded regions show $68\%$ contours for the expected distribution of a data-like sub-sample from the SMALL simulation.
}
\label{fig:v90feedz3}
\end{figure*}

Figure \ref{fig:v90feedz3} shows the velocity width probability distribution function at $z=3$, comparing each of our simulations to observations.
\footnote{We have normalized the pdf so that the integral over the full range is unity; this differs 
from \cite{Pontzen:2008} and \cite{Tescari:2009}, who show the velocity width PDF normalized by the line density per unit absorption 
distance of DLAs.} The observed sample of DLA velocity widths in N13 does not show significant redshift evolution.
The redshift evolution in our simulations is also mild, although there are $40\%$ more spectra with velocity widths $< 18$ \kms~at $z=4$ than $z=3$, 
due to a lower amplitude of the photo-ionizing background. We therefore compare the combined observational catalogue for
$z=1.7-5.1$ to our simulations at $z=3$, the mean redshift of the catalogue. As discussed in N13, the velocity width distribution shows signs of 
two peaks; one at $30$ \kms, and one at $100$ \kms. However, our error analysis confirms the conclusion of N13 that this is consistent with 
a single-peaked underlying model.


\begin{figure*}
\includegraphics[width=0.33\textwidth]{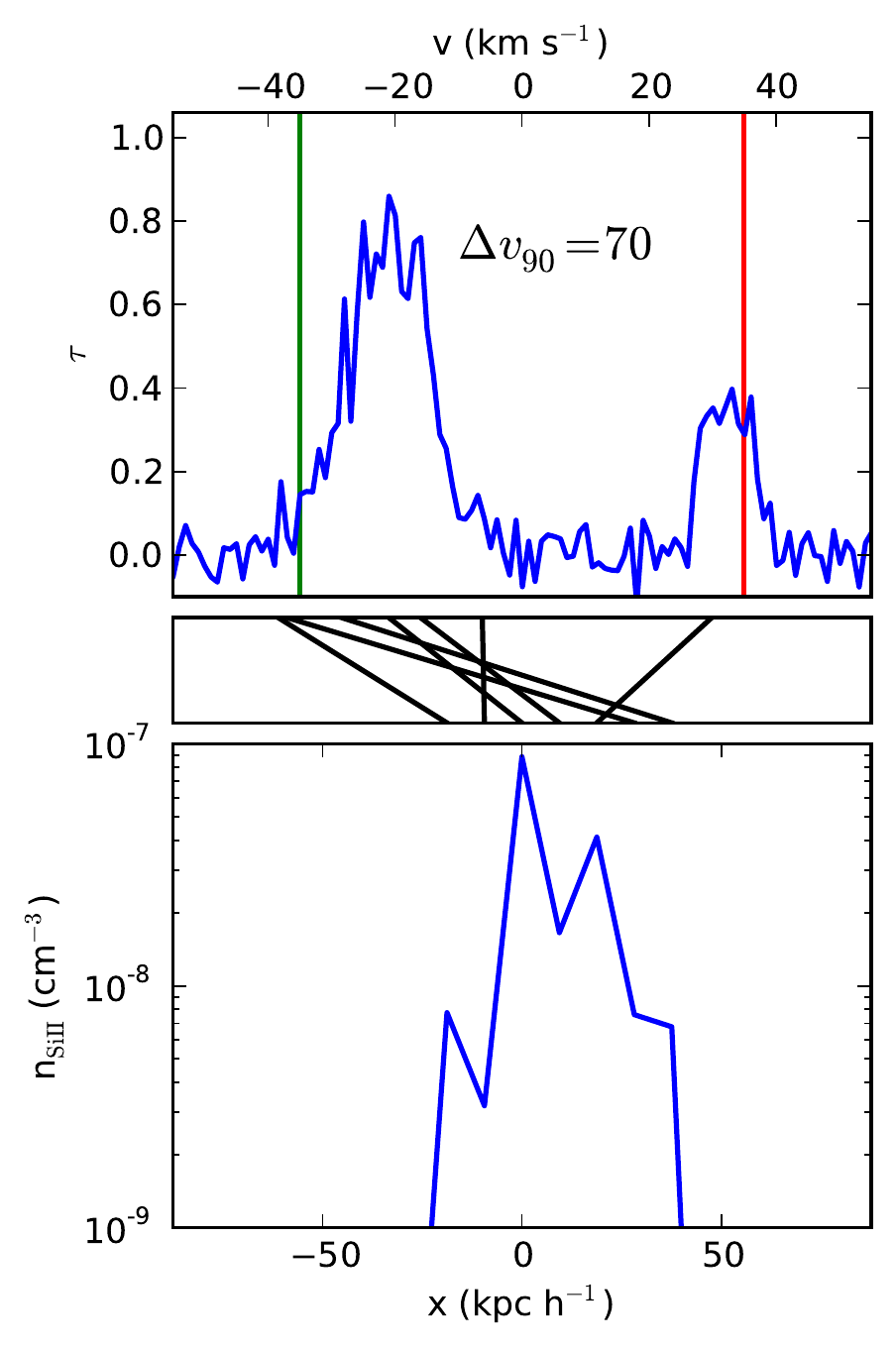}
\includegraphics[width=0.33\textwidth]{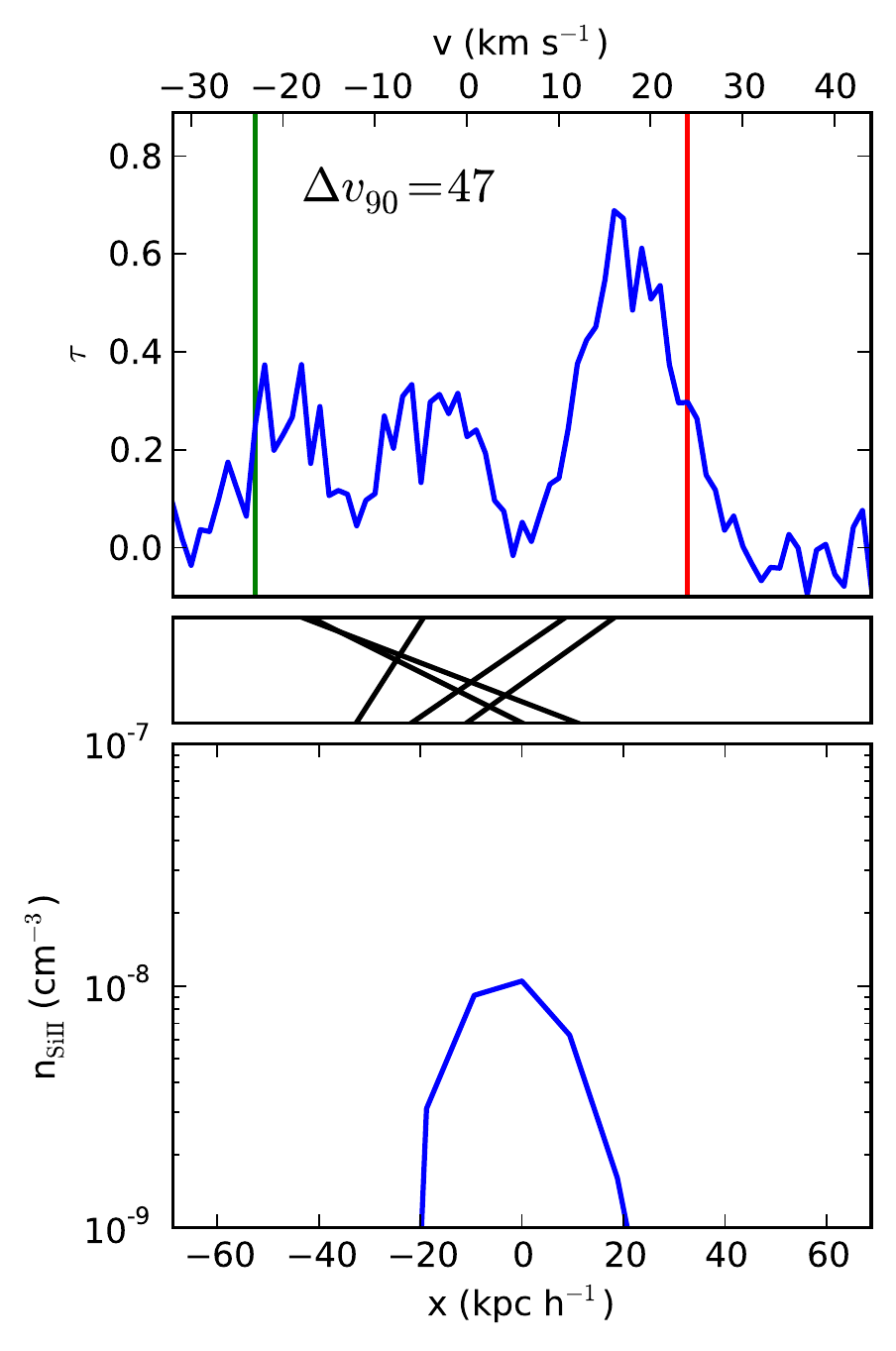}
\includegraphics[width=0.33\textwidth]{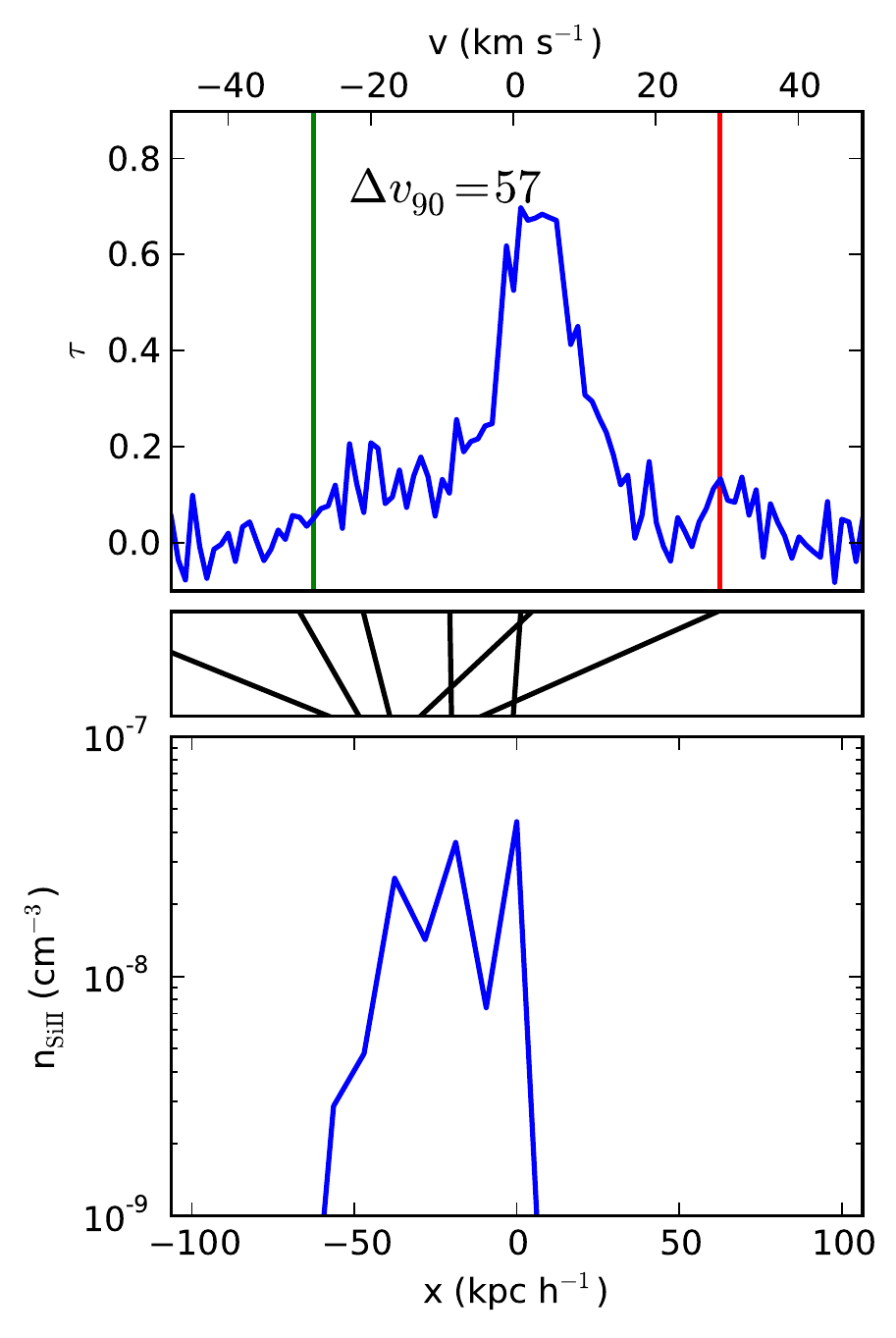} \\
\caption{Top: example \SiII~spectra with typical velocity widths from a $10 \Mpch$ box at $z=3$. 
Gaussian noise has been added for easier comparison to observations, assuming a SNR of $20$.
Vertical green and red lines show the bounds of the velocity width.
Bottom: corresponding \SiII~density for each spectrum.
Velocities are in \kms, and physical positions are in comoving $\kpch$.
At this redshift $1\,\mathrm{km s}^{-1} = 9.5  \kpch$. 
Central: correspondence between physical density and velocity and thus the effect of peculiar velocity.
}
\label{fig:smallv90spectra}
\end{figure*}

The DEF simulation is in generally good agreement with the data, marking the first time the velocity width distribution 
has been reproduced with such fidelity. The main reasons for this, discussed further in Appendix \ref{ap:compare}, are our 
strong feedback and our treatment of the \SiII~fraction.

Figure \ref{fig:smallv90spectra} shows three example spectra, together with the
\SiII~density along the sightline. The HI density has a similar shape, but smaller secondary peaks,
due to \SiII~existing at lower gas densities (see Figure \ref{fig:ionfrac}).
The spectra have been plotted including Gaussian noise with a SNR of $20$, to allow easy by-eye comparison to the observed spectra shown in N13.
We have used the higher resolution simulation, SMALL, because the small-scale spectral structure was not entirely converged in the larger box. 
The lower resolution caused reduced velocity broadening, and absorption profiles 
overly concentrated into peaks, with little absorption between them. A by-eye comparison 
of the simulated spectra to those observed indicates that even in our highest resolution simulations the simulated spectra may still have too peaked a shape,
perhaps indicating that a further increase in resolution or a more detailed model of the interstellar medium is needed. We have included the results from SMALL in Figure \ref{fig:v90feedz3}, 
to show the effect on the velocity width distribution, which is modest. The main difference is a reduced population of 
small velocity width systems, which are those most affected by the broadening, and the corresponding increase in systems with a velocity width of $\sim 100$ \kms.

A few small discrepancies remain between the simulation data and the observations. As revealed by the cumulative velocity width distribution (Figure \ref{fig:v90feedz3}), 
the DEF simulation shows a distinct preference for smaller velocity width systems. This is somewhat ameliorated by increased resolution, 
but some tension remains at the $20\%$ level. Models where the characteristic DLA halo mass is larger than in our simulations 
are thus certainly not ruled out, and a moderately larger DLA host halo mass may be favoured. 

The NOSN and HVEL simulations, which produce sufficient spectra in the tail of the velocity width distribution, show a peak at a significantly lower velocity width than the observations. 
The observed velocity width distribution (and DEF) peak at $60-80$ \kms, while NOSN and HVEL peak at $30$ \kms. The FAST simulation produces a peak at $40-60$\kms, 
intermediate between DEF and NOSN, and in $1-\sigma$ tension with the observations. Note that this simulation performed relatively well in \cite{Bird:2014}, 
demonstrating the extra constraint on feedback processes from the velocity width distribution. 

\begin{figure}
\includegraphics[width=0.45\textwidth]{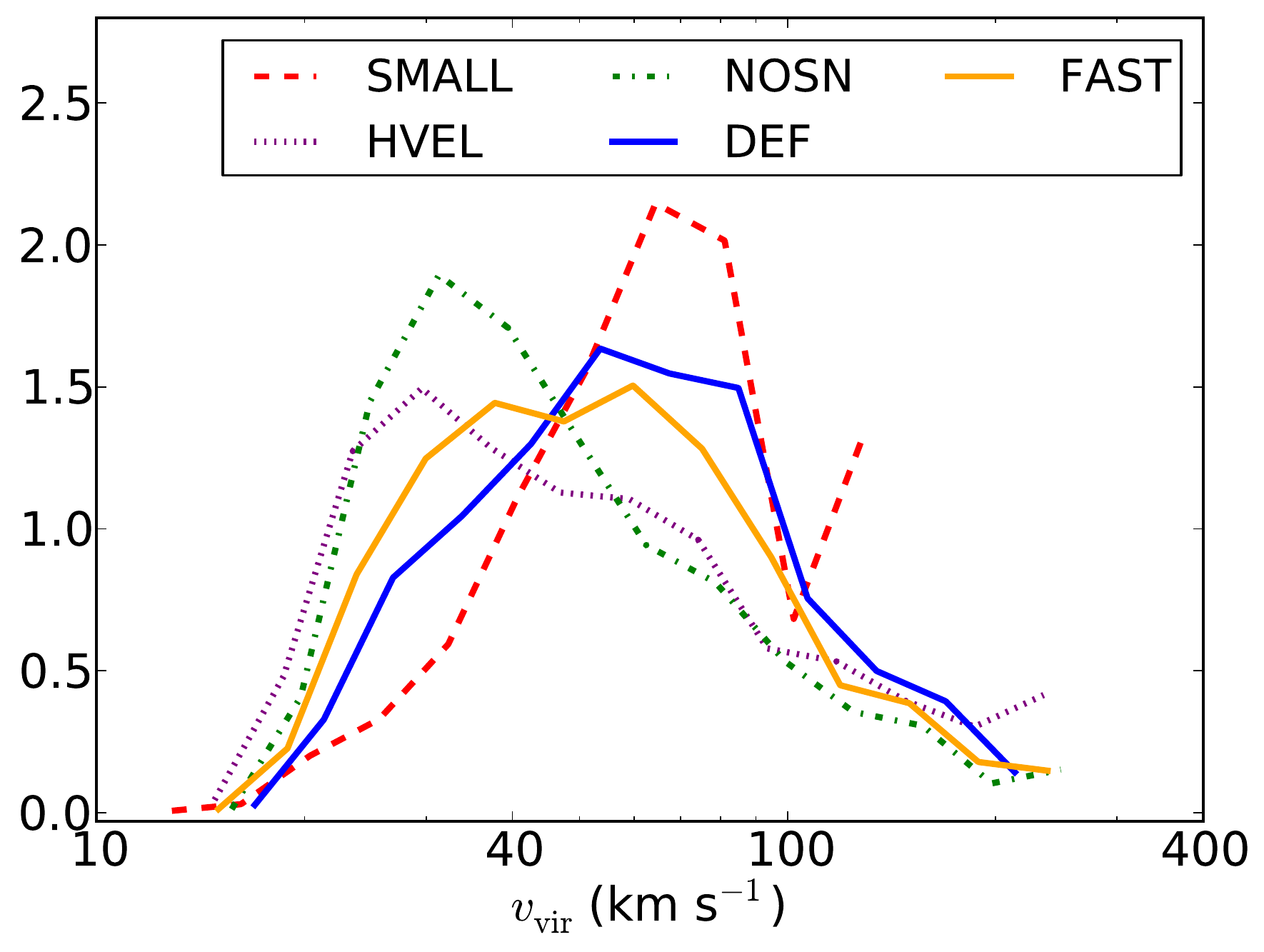}
\caption{The distribution of virial velocities of the haloes hosting DLAs in our simulations.}
\label{fig:v90vvir}
\end{figure}

\begin{figure}
\includegraphics[width=0.45\textwidth]{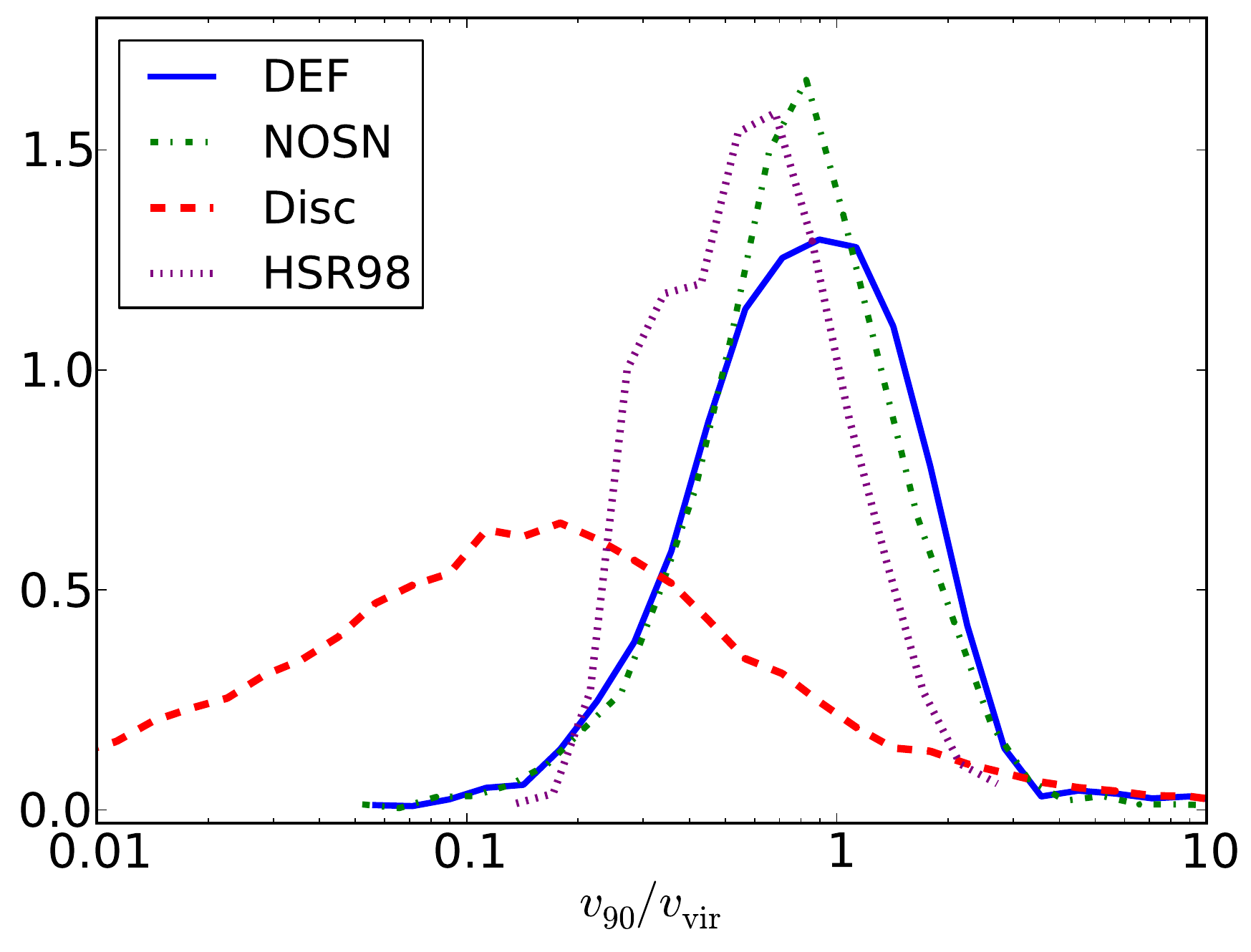}
\caption{Histograms of the ratio of velocity width to halo virial velocity at $z=3$, $\velwm / \vvir$, for our 
simulations. FAST is similar to DEF, and HVEL to NOSN. We compare to the rotating disc model described in Appendix \protect\ref{ap:disc} (Disc) and 
the simulations of \protect\cite{Haehnelt:1998}.
}
\label{fig:v90halomass}
\end{figure}

Figure \ref{fig:v90vvir} shows the virial velocity, $v_\mathrm{vir} = \sqrt{G M / R_{200}}$, of DLA host 
haloes.\footnote{$R_{200}$ is the radius within which the average density is $200$ times the critical density.}
In DEF and FAST the characteristic virial velocity is $v_\mathrm{vir} \sim 70$ \kms, while, in HVEL and 
NOSN, DLAs have $v_\mathrm{vir} \sim 30$ \kms. In the latter simulations, feedback is less effective at suppressing DLAs in small haloes. 
Note that the $10 \Mpch$ box used in SMALL is too small to include a representative sample of the largest haloes hosting DLAs.
Figure \ref{fig:v90halomass} shows the ratio between virial velocity and velocity width. 
In all simulations this ratio is of order unity, demonstrating that the velocity width does indeed
probe the virial velocity of the DLA host halo, albeit with a large scatter.
The NOSN and HVEL simulations have $\velwm \sim 0.7 v_\mathrm{vir} $, 
while the stronger feedback in DEF and FAST produces $\velwm \sim 0.9 v_\mathrm{vir}$. 
However, the larger change in the characteristic virial velocity of DLAs shown in Figure \ref{fig:v90vvir} is more important for explaining 
the differences between the velocity width distribution. 

\subsubsection{Disc Models}
\label{sec:discs}

Figure \ref{fig:v90halomass} compares our simulations to the rotating disc model described in Appendix \ref{ap:disc} and the results of \cite{Haehnelt:1998}.
The latter, also based on simulations without feedback, are remarkably similar to our NOSN model. The differences are probably due to our lower self-shielding threshold.
The disc model shows the expected relationship between velocity width and virial velocity for a sight-line intersecting a thin disc 
at a random angle. It predicts a much broader velocity width distribution peaking at a lower fraction of the virial velocity. 
The location of the peak is controlled by the ratio of the disc scale height to scale length, $r_l$. Thicker discs produce a larger velocity width, 
as they maximize the difference between the radius at which the sight-line enters the disc and the radius at which it exits. 
As explained in Appendix \ref{ap:disc}, we have plotted the distribution for $r_l = 0.25$.
Even in this case, the disc model is not a good match to our simulations. We checked more explicitly for rotation by finding the fraction of 
gas cells in our simulation whose position within a halo is rotationally supported. 
To do this, we computed the velocities parallel and perpendicular to the vector from each cell to the halo centre, $v_\mathrm{par}$ and $v_\mathrm{perp}$.
We required that cells be within the halo virial radius, have a gas density greater than $3 \times 10^{-4}$~cm$^{-3}$ and 
move azimuthally, with $| v_\mathrm{perp}|  > 2 |v_\mathrm{par}|$. We then computed the 
halo rotational velocity at the cell radius, $v_\mathrm{rot}$, assuming a Navarro-Frenk-White profile \citep{NFW} with a concentration of $10$, as for a Milky Way-sized halo, 
and required $0.7 < v_\mathrm{perp} / v_\mathrm{rot} < 1.5$.
Note that we are checking for rotational support, not the presence of a disc; for the latter we should also check that the angular 
momentum vectors of the cells are close to the mean angular momentum vector. 
We find that in the DEF simulation less than $10\%$ of the considered cells were rotationally 
supported within their host halo ($5\%$ in the HVEL simulation), 
and the velocity width distribution of the absorption spectra resulting from these cells peaked at $\sim 20$\kms.

\subsubsection{Large Velocity Widths}
\label{sec:largevel}

\begin{figure*}
\includegraphics[width=0.33\textwidth]{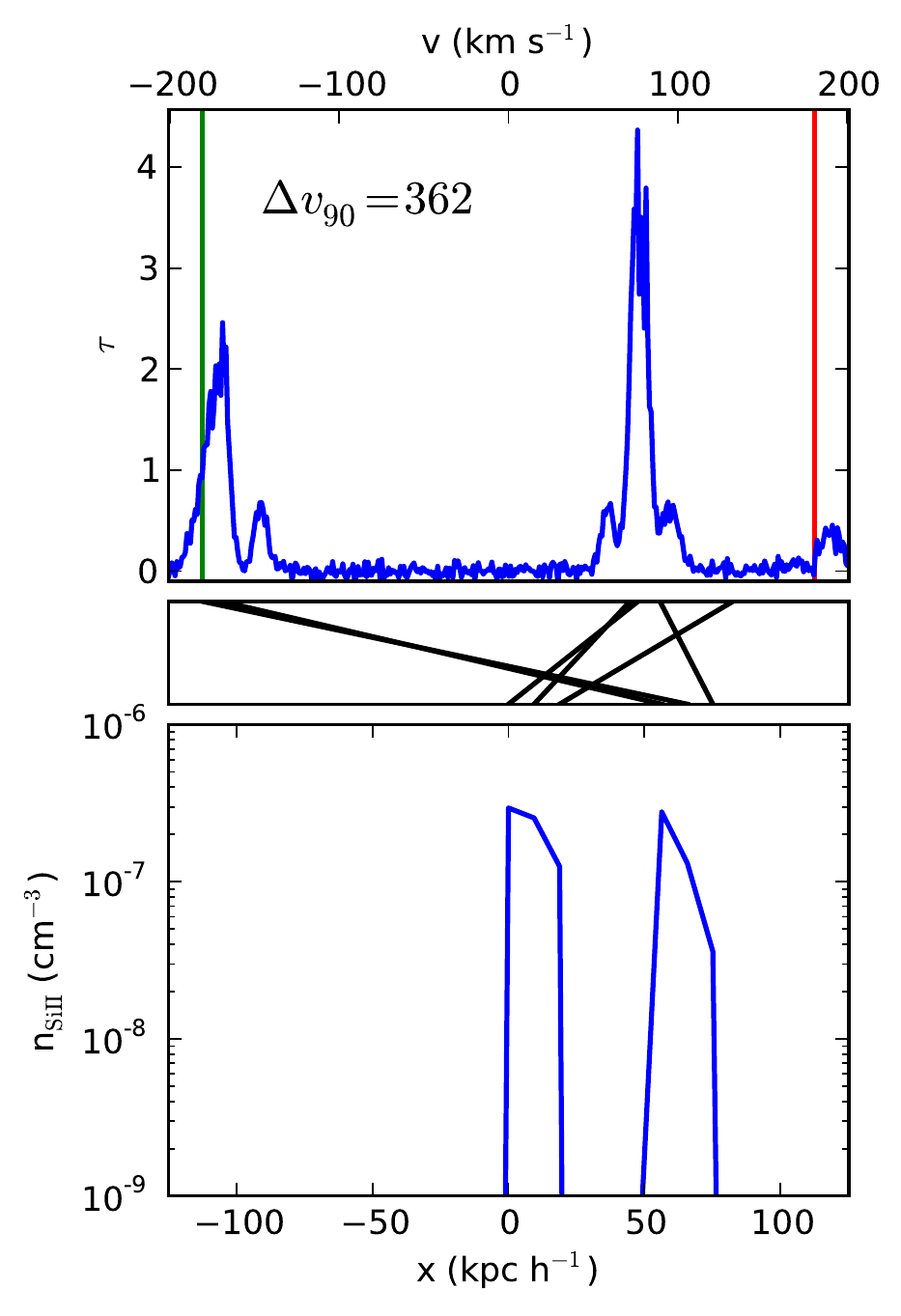}
\includegraphics[width=0.33\textwidth]{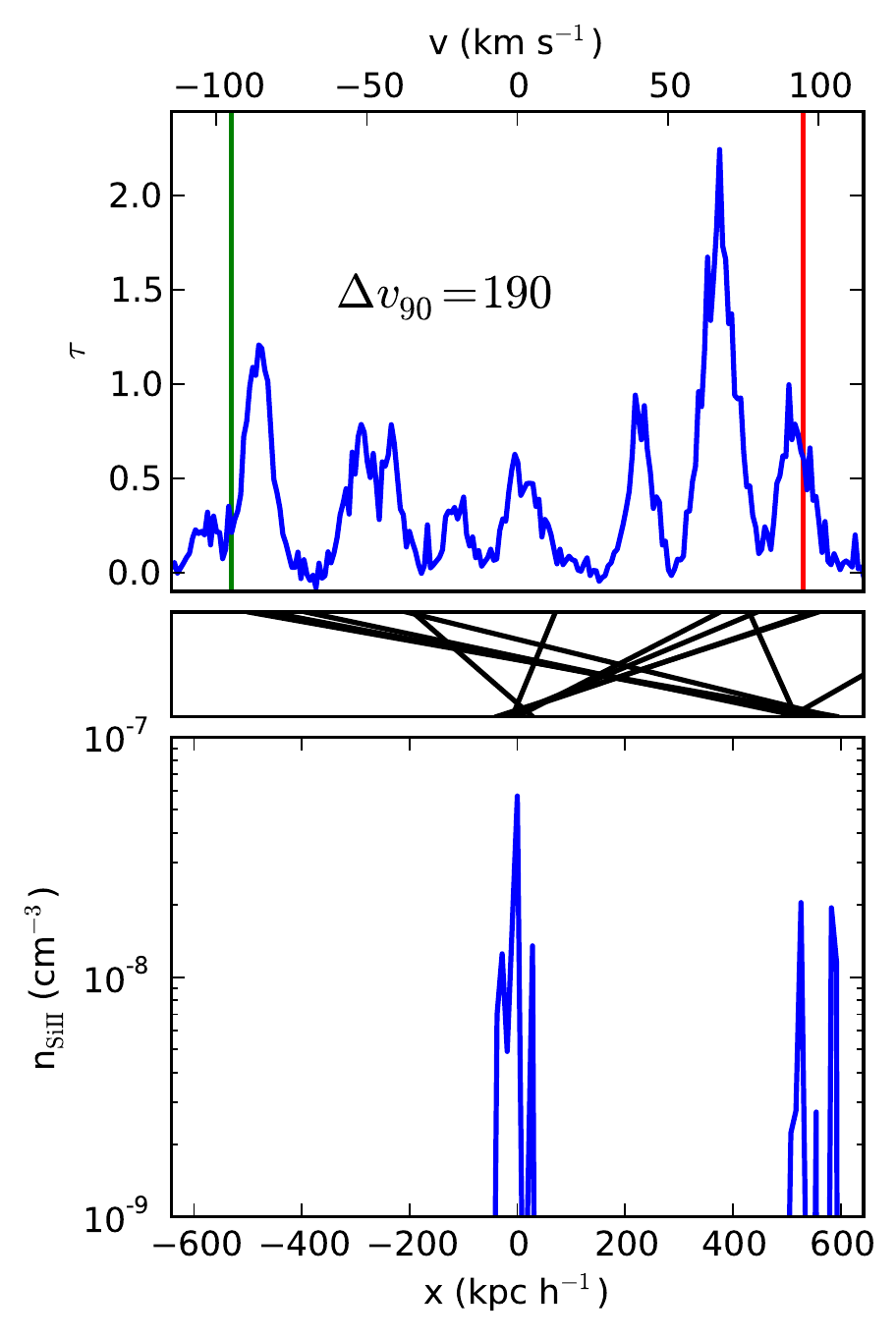}
\includegraphics[width=0.33\textwidth]{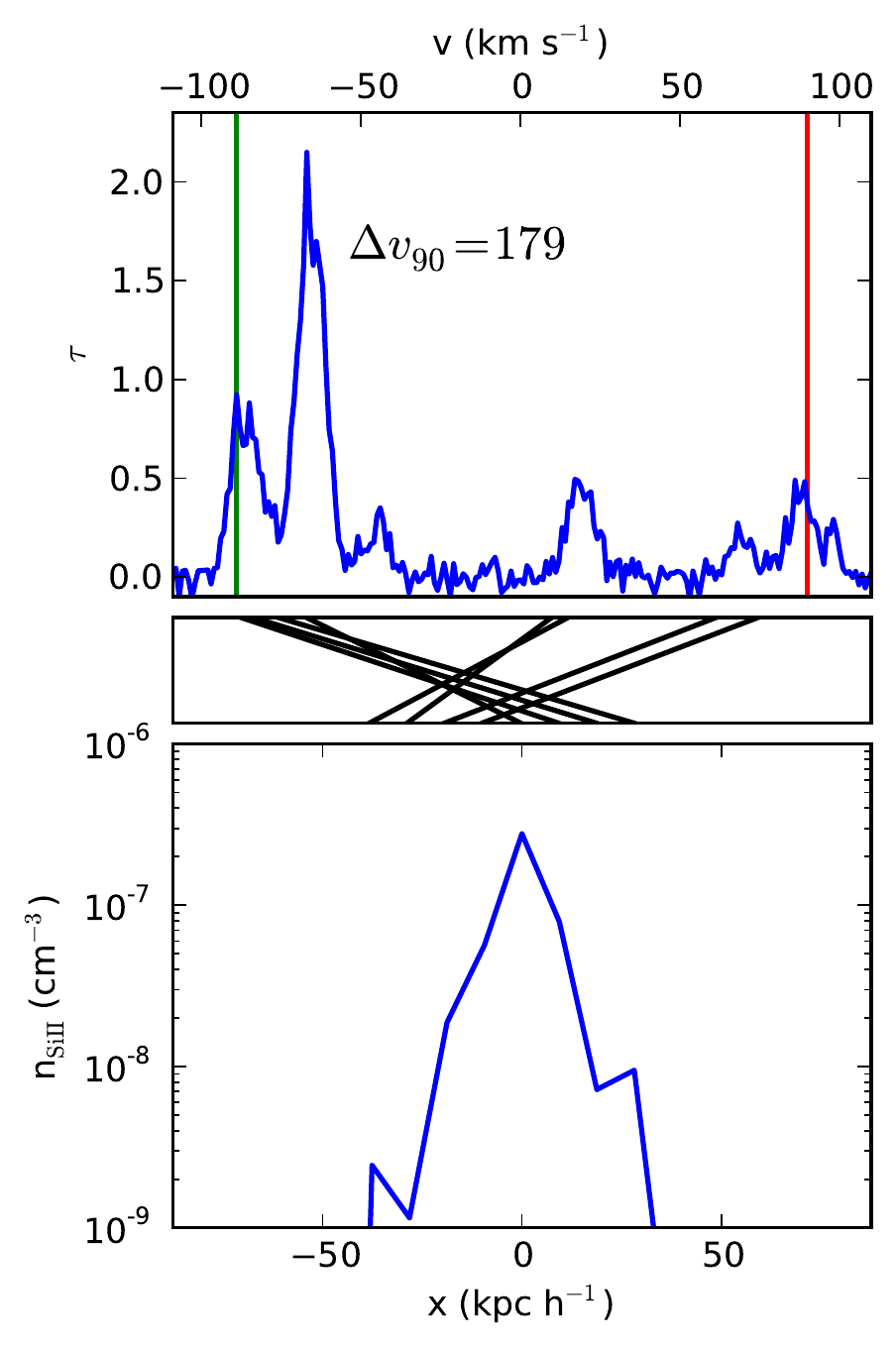} \\
\caption{(Top) Example \SiII~spectra with velocity widths $> 150$ \kms, from a $10 \Mpch$~simulation at $z=3$.
Gaussian noise has been added for easier comparison to observations, assuming a SNR of $20$.
Vertical green and red lines show the bounds of the velocity width.
(Bottom) Corresponding \SiII~density for each spectrum.
Velocities are in \kms, and physical positions are in comoving $\kpch$.
At this redshift $1\, \mathrm{km s}^{-1} = 9.5  \kpch$. 
(Central) Correspondence between physical density and velocity and thus the effect of peculiar velocity.
}
\label{fig:largev90spectra}
\end{figure*}

The observational data contains velocity widths up to $\sim 400$ \kms, which some earlier simulations have had difficulty reproducing.
The largest halo virial velocity in our simulations is around $200$ \kms, while the largest velocity width is $\sim 600$ \kms; 
there is thus a population of objects with velocity widths that significantly exceeded the virial velocity of their host halo. 

\begin{figure}
\includegraphics[width=0.45\textwidth]{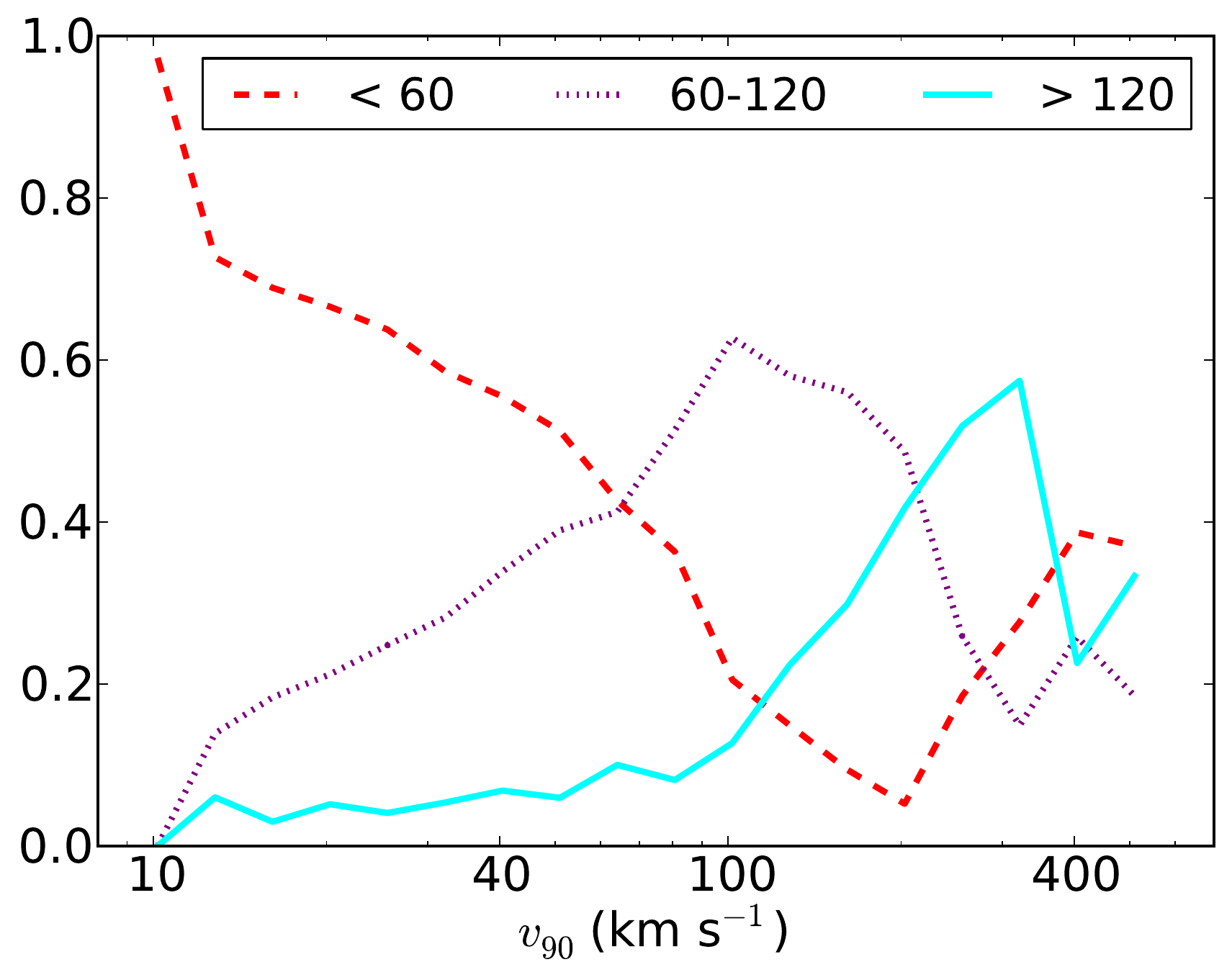}
\caption{Fraction of spectra from the DEF simulation at $z=3$ arising from haloes in one of three mass ranges, as a function of velocity width.}
\label{fig:break}
\end{figure}

Figure \ref{fig:largev90spectra} shows examples of these spectra.
All exhibit significant internal structure and multiple absorption peaks.
We show examples of absorption profiles arising from both spatially extended objects and spatially compact but velocity broadened objects. 
Ninety percent of sightlines with $\velwm < 100$ \kms~are spatially compact, but this changes for larger velocity widths; half of the sightlines 
with $\velwm = 250$ \kms~show multiple peaks in density. This fraction is consistent with that in the observational sample N13.
The majority of these systems are associated with a DLA and an LLS aligned along a sight-line, with 
the LLS residing in a filament in the IGM.
Figure \ref{fig:break} emphasises this point, showing the host halo virial velocity for spectra as a function of velocity width.
For $\velwm < 200$ \kms, the velocity width correlates well with the virial velocity of the halo, as expected.
However, for $\velwm \geq 200$ \kms~the fraction of spectra coming from the smallest virial velocity bin starts to increase, reflecting
the increasing fraction of multiple density peak systems arising from a small DLA-hosting halo and an LLS.

\begin{figure}
\includegraphics[width=0.45\textwidth]{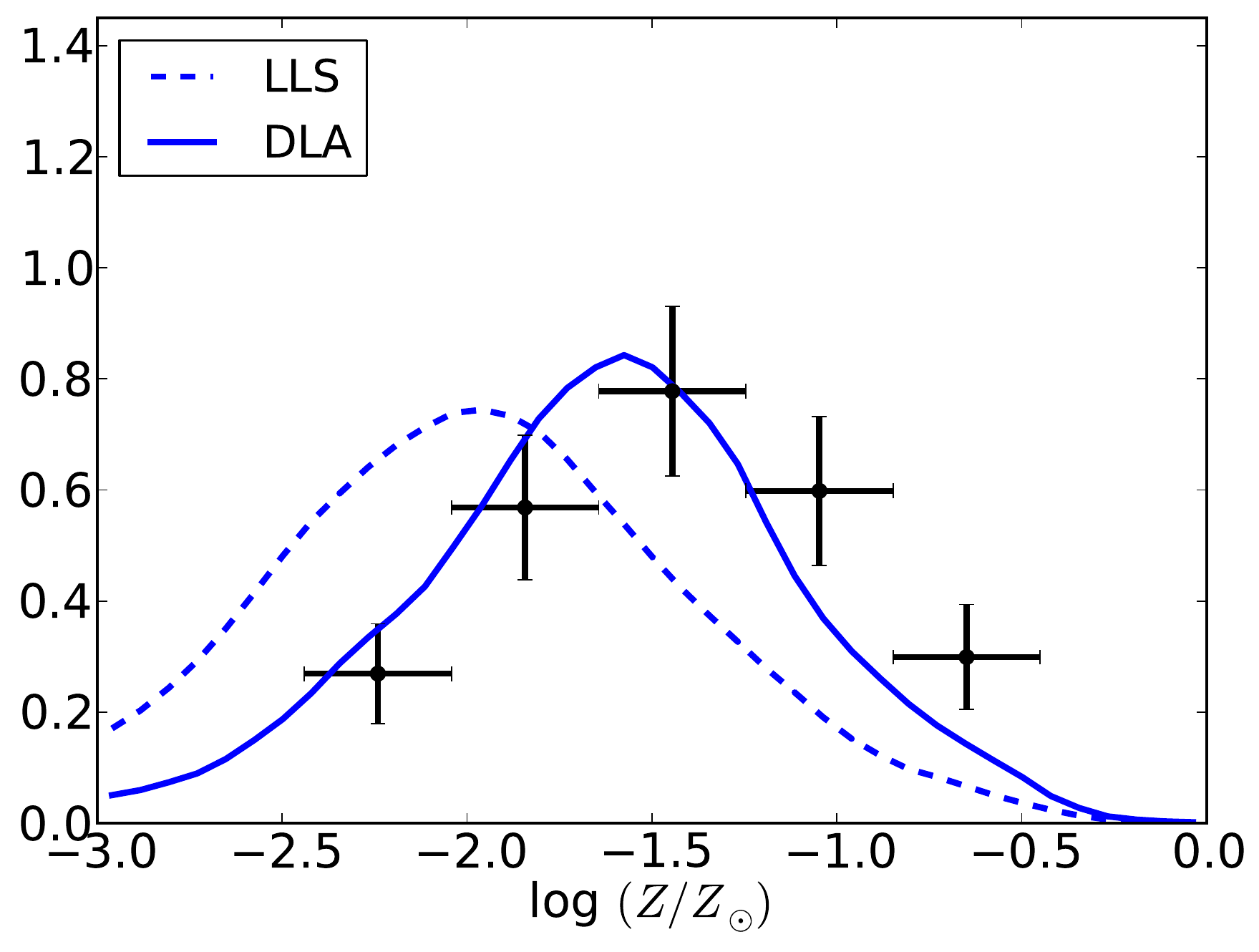}
\caption{Metallicity distribution of DLAs (solid) and LLSs (dashed) from the DEF simulation at $z=3$.
Both were computed as described in \protect\cite{Bird:2014}, and are shown as a fraction of the solar metallicity, 
which we take to be $M_{Z_\odot} / M_{H_\odot} = 0.0134/0.7381$ \protect\citep{Asplund:2009}.
Data points show DLA metallicity from \protect\cite{Rafelski:2012}, assuming Poisson errors. }
\label{fig:llsmetals}
\end{figure}

These statistics imply the existence of a population of metal-enriched LLSs, capable of producing an 
absorption trough containing at least $10$\% of the total optical depth.
Figure \ref{fig:llsmetals} shows the metallicity distribution of DLAs and LLSs for the DEF simulation.
LLSs are relatively highly enriched, with a mean metallicity $\sim 60$\% of the mean DLA metallicity. 
The stellar feedback model in DEF drives strong, metal enriched, outflows, and a significant fraction of the LLS enrichment arises
directly from them. Note that this makes the LLS metallicity sensitive to parameters which do not affect the DLA metallicity, 
such as the wind metal loading. As the NOSN simulation lacks outflows, only a small fraction of metals escape star-forming regions,
but the star formation rate is sufficiently high that LLSs are still enriched.

%

\subsection{Edge-Leading Spectra}
\label{sec:fmeanmedian}

\begin{figure*}
\includegraphics[width=0.45\textwidth]{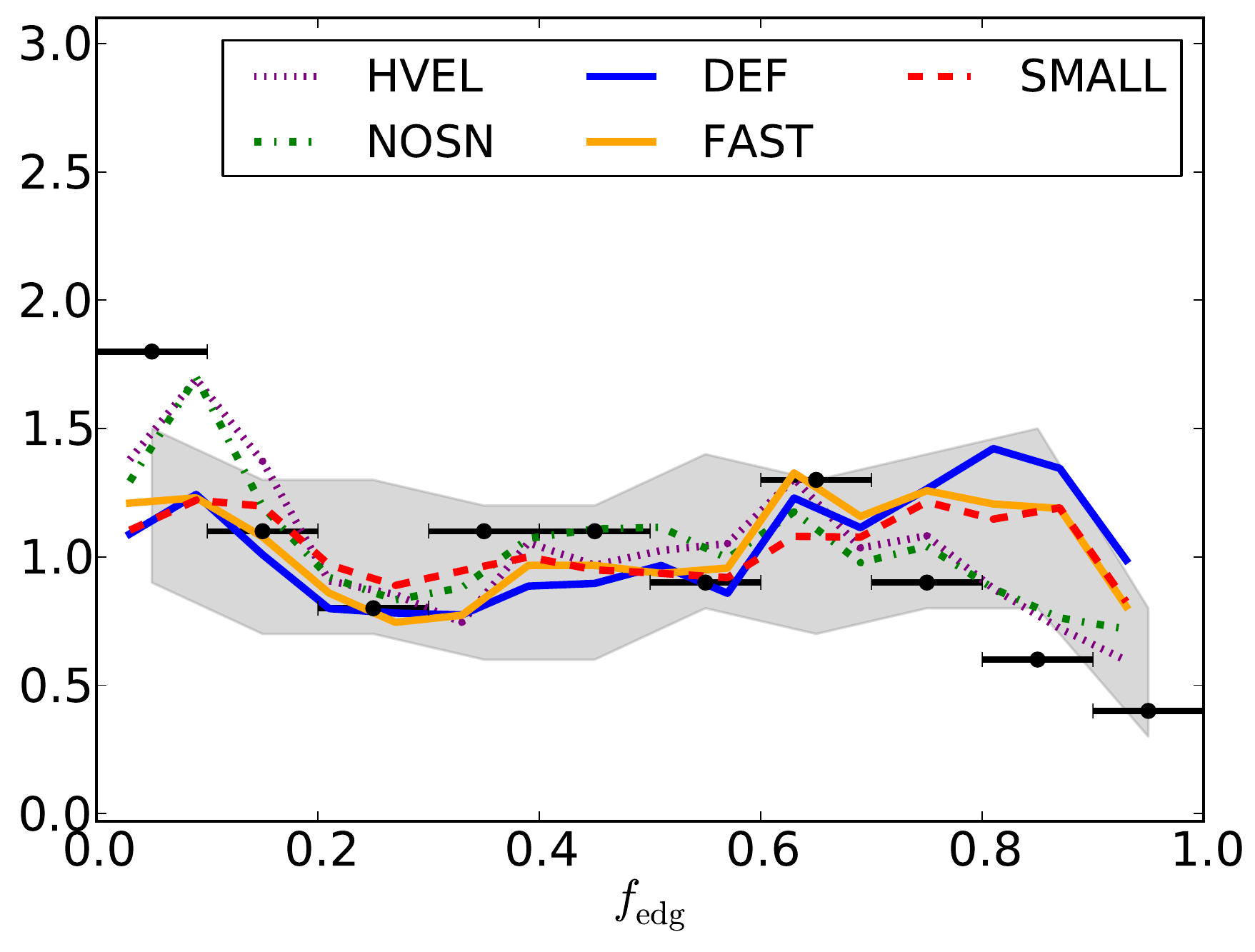}
\includegraphics[width=0.45\textwidth]{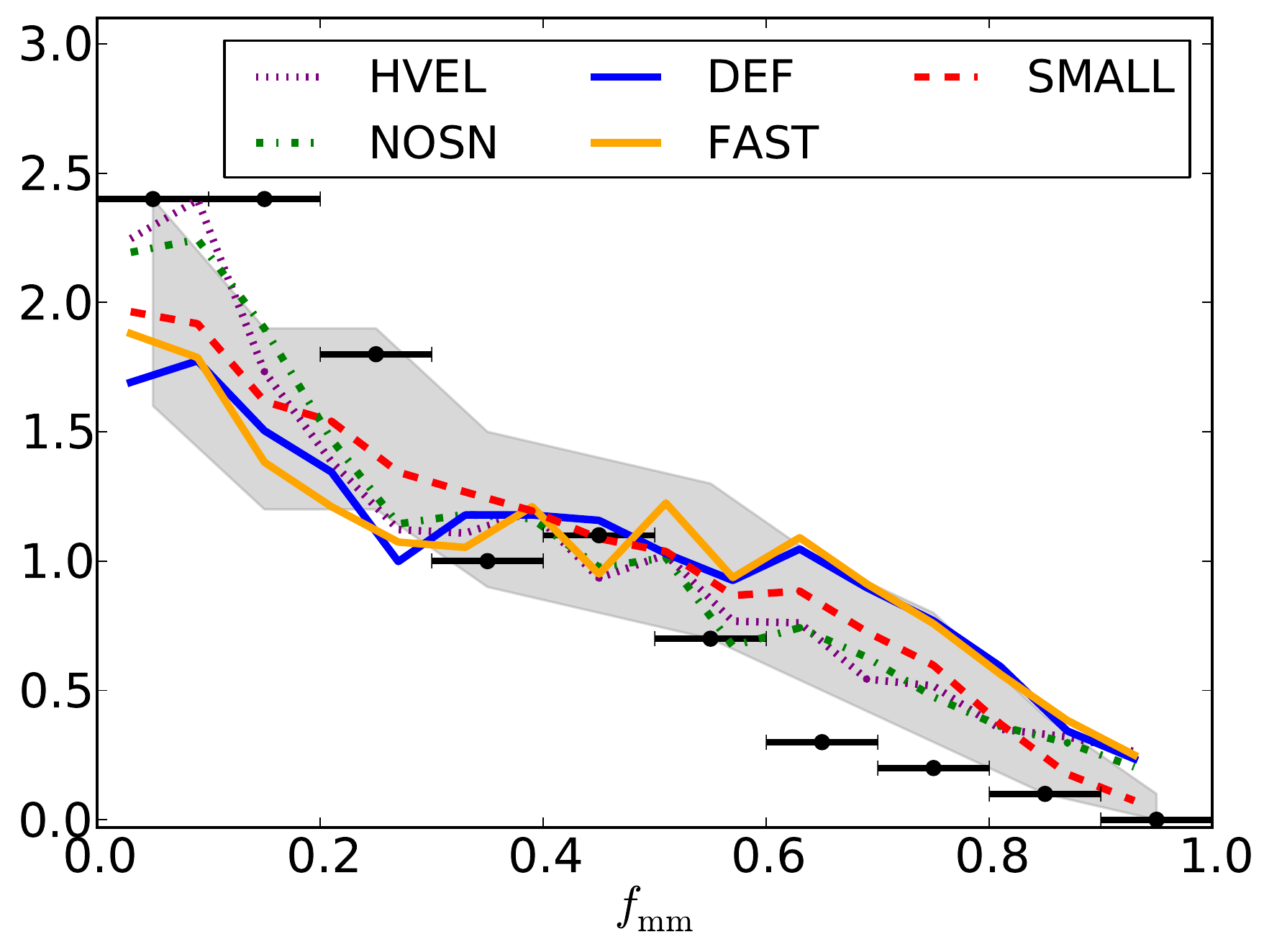}
\caption{
Probability density functions at $z=3$ for (left) the edge-leading statistic and (right) the mean-median statistic compared to 
that computed from the spectra of N13. Grey shaded regions show $68\%$ contours for the expected distribution of 
a data-like sub-sample from the SMALL simulation.
}
\label{fig:fmmedge}
\end{figure*}

Figure \ref{fig:fmmedge} shows the distribution of $\fedg$ and $\fmm$, the edge leading and mean median statistics defined in Section \ref{sec:kinematics}.
As with the velocity width distribution there is no observable redshift evolution, so once again we show the observational results
at all redshifts compared to the simulation output at $z=3$. 

Convergence with resolution in $\fedg$ is good, although the abundance of very edge-leading 
spectra is slightly lower in the higher resolution simulation, due to the shift in the velocity 
width distribution. The distribution of $\fmm$, on the other hand, is more affected. 
Higher resolution broadens secondary absorption peaks. This does not affect $\fedg$, which is only sensitive
to the strongest absorption, but it tends to smooth the overall spectrum and thus reduce $\fmm$.

All simulations produce too many spectra with a large value of $\fmm$, with the discrepancy worse for 
DEF and FAST. FAST and DEF also have a stronger preference for edge-leading spectra than NOSN and HVEL, 
producing as many spectra at $\fedg = 0.8$ as at $\fedg = 0.6$, while the observations, HVEL, and 
NOSN show a drop.\footnote{Note that $\fedg$ is not strongly correlated with \velw, so large values of $\fedg$ do not necessarily correspond to 
spectra with large velocity widths. The population of spectra with large $\fedg$ is dominated by systems 
resembling the middle and left-hand panels of Figure \ref{fig:smallv90spectra}, and so is not affected by any potential 
ambiguity in absorber size.} By eye, this excess seems like a significant failure of the model. However, our bootstrap error 
analysis shows that at present the sample is too small to draw a definitive conclusion and the observed data are
still marginally consistent with the simulation.
If the discrepancy is real, it has important implications for the preferred feedback model.
In our simulations, star formation in haloes with a small virial velocity is suppressed by outflows. This gives gas 
a relatively large radial velocity with respect to the halo and thus produces a high $\fedg$.
If there were also some other method, such as thermal heating, for suppressing star formation, $\fedg$ could be lower while 
still producing enough large velocity width systems. 

\subsection{SiII~1526 Equivalent Width}
\label{sec:eqwsiii}

\begin{figure*}
\includegraphics[width=0.33\textwidth]{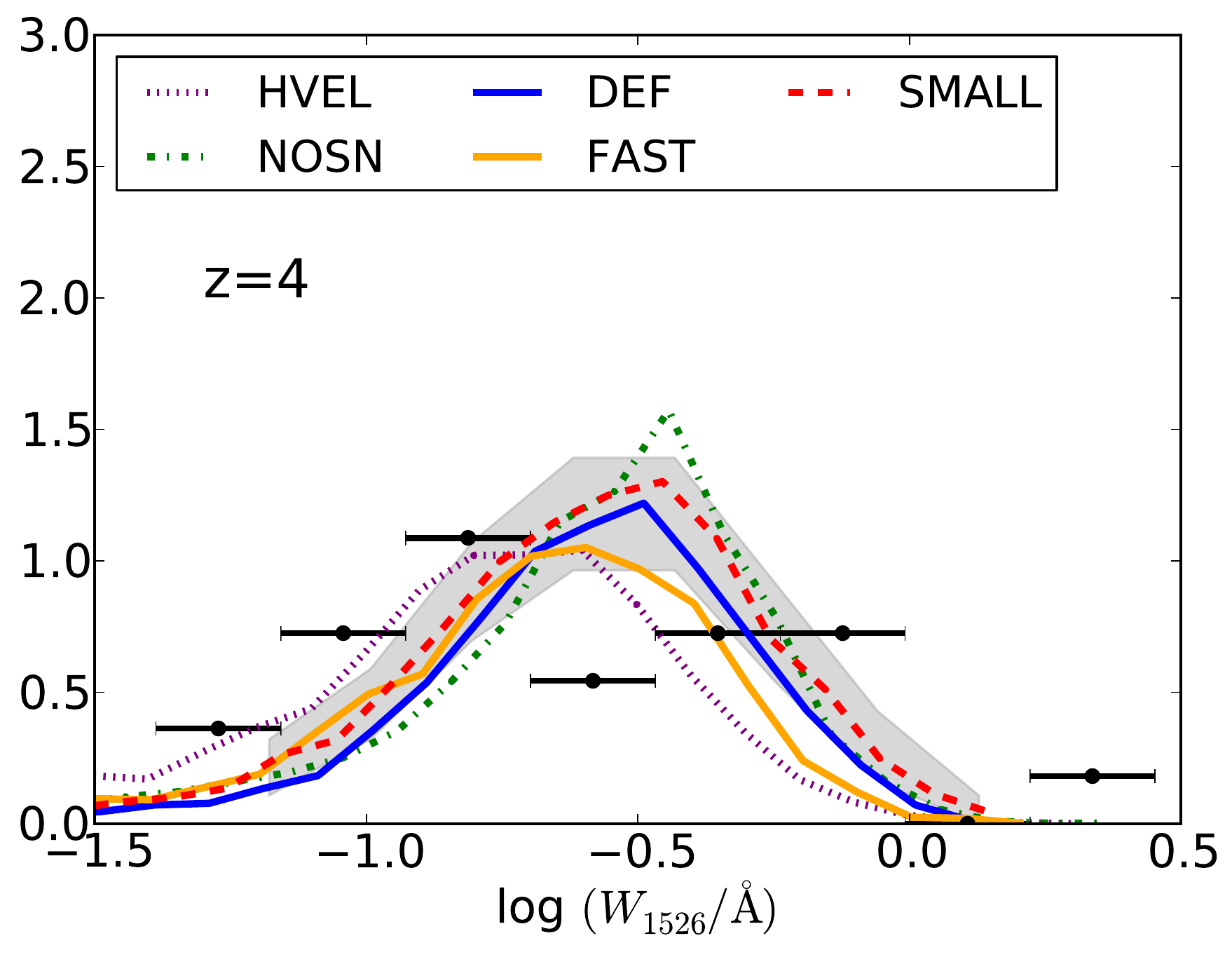}
\includegraphics[width=0.33\textwidth]{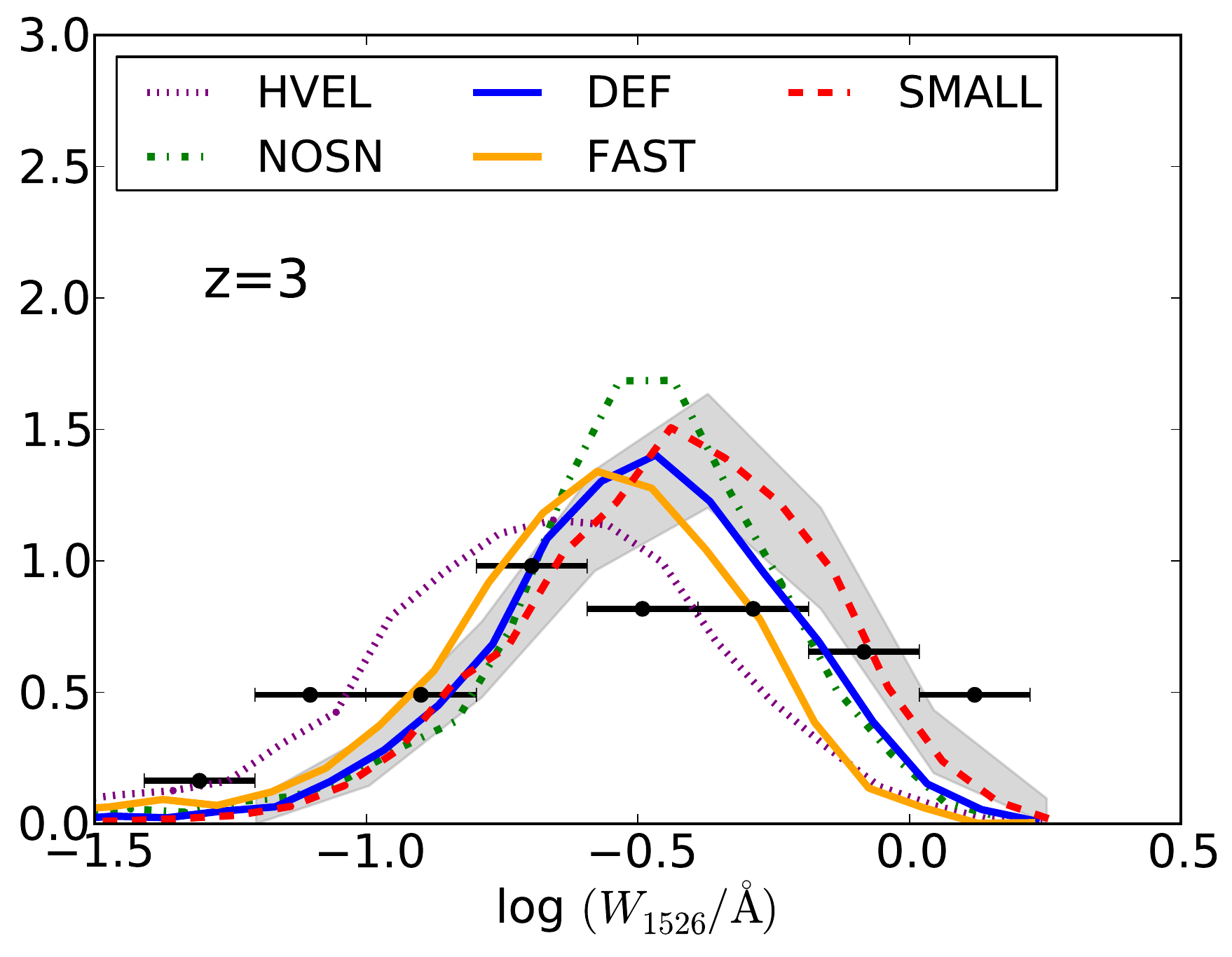}
\includegraphics[width=0.33\textwidth]{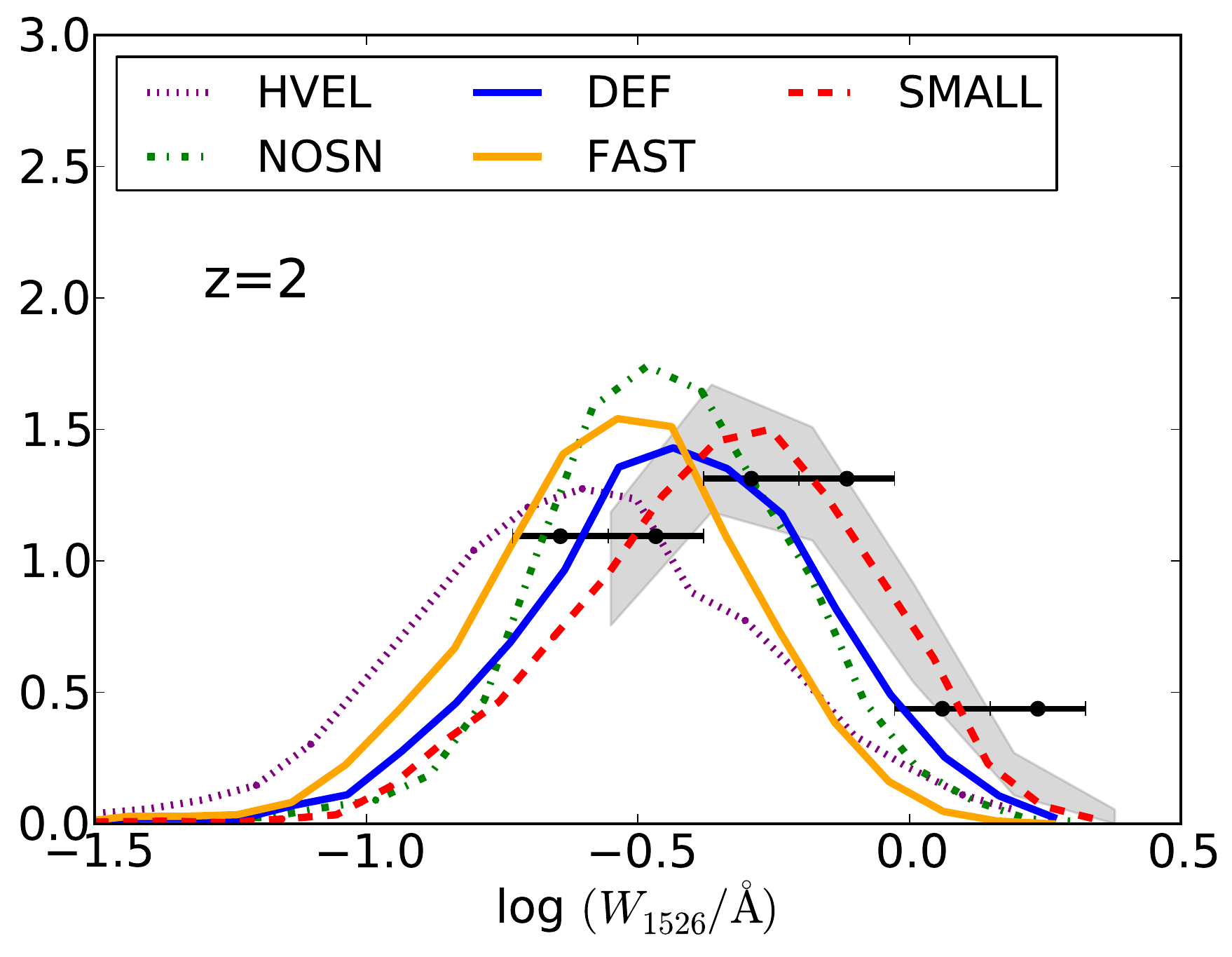}
\caption{The equivalent width distribution of \SiII~$1526$ \AA~for each of our simulations at (left) $z=4$, (center) $z=3$ and (right) $z=2$, 
compared to N13. Grey shaded regions show $68\%$ contours for the expected distribution of 
a data-like sub-sample from the DEF simulation.
}
\label{fig:siiieq}
\end{figure*}

Figure \ref{fig:siiieq} shows the equivalent width distribution for the \SiII~$1526$ \AA~line, defined as
\begin{equation}
 W_\mathrm{1526} = \int 1 - \exp(-\tau) d\lambda\,.
\end{equation}
The equivalent width is sensitive to the total amount of the ionic species. Thus, 
unlike our other results, it depends on our self-shielding prescription; turning 
off self-shielding for the \SiII~decreases the equivalent width by approximately $25$\%. 
However, because the \SiII~$1526$ \AA~line is usually saturated, it depends only weakly on the total 
metallicity, especially in the denser regions.

All our simulated results are a reasonable match to the observations, reproducing the mean and standard deviation 
of the equivalent width distribution, although the data prefer a distribution which is more extended at the large 
equivalent width end. Convergence is very good at $z=4$, but at lower redshifts SMALL produces 
moderately larger equivalent widths, due to a slight increase in star formation at higher resolution.

The variation in the simulations mirrors the variation in the DLA metallicity \citep{Bird:2014},
but because the line is saturated the distinctions between them are partially erased. In particular the NOSN simulation produces 
a much larger DLA metallicity than the others, but a similar equivalent width distribution, because the high density regions 
which contain most of the extra metals give rise to saturated absorption.

We therefore caution against over-interpreting the equivalent width results. 
The velocity width is insensitive to the total amount of \SiII~present and thus particularly robust. 
By contrast, the equivalent width is sensitive to the amount of \SiII~at relatively low densities. We found that
it was affected relatively strongly by our treatment of self-shielding, which assumes the self-shielding factor depends only on gas
density and temperature. We have neglected any explicit dependence on the density of the surrounding gas, which 
may be the cause of the wider scatter in the observations.

\subsection{Velocity Width Metallicity Correlation}
\label{sec:correlation}


A correlation is observed between velocity width and metallicity \citep{Ledoux:2006, Moller:2013, Neeleman:2013},
analogous to the mass-metallicity relation for galaxies \citep{Erb:2006}. As we showed in Section \ref{sec:velwidth}, velocity width 
generally traces the DLA host virial velocity, and thus provides an analogue to the halo mass. The correlation is relatively
weak because, as shown in Figure \ref{fig:v90halomass}, there is significant scatter in the relation between velocity width and halo virial velocity.

\begin{table}
\begin{center}
\begin{tabular}{|c|c|c|c|c|c|}
\hline
Data set &   $z$  	  &   KS test   &P(KS test)&  Pearson $r$ \\
\hline 
DEF 	&  $4$	  &   $0.2$	& $0.4$		&  $0.4$	\\
DEF 	&  $3$	  &   $0.2$	& $0.3$		&  $0.5$	\\
DEF 	&  $2$	  &   $0.3$	& $0.04$	&  $0.46$	\\
HVEL 	&  $3$	  &   $0.4$	& $0$		&  $0.36$	\\
NOSN 	&  $3$	  &   $0.6$	& $0$		&  $0.33$	\\
FAST 	&  $3$	  &   $0.3$	& $0.06$		&  $0.3$ \\
Observed &	$5-3.5$	  &   -	& -			&  $0.6$ \\
Observed &    $3.5-2.5$	  &   -	& -			&  $0.6$ \\
Observed &    $2.5-1.7$	  &   -	& -			&  $0.81$ \\
\hline
\end{tabular}
\end{center} 
\caption{Table of Statistical tests.
KS test shows the result of a 2D KS test between the shown simulation output and the observations at that redshift. 
KS test probability is calibrated by taking random sub-samples of the simulated data and comparing them to the full sample.
Pearson $r$ shows the correlation between metallicity and velocity width.
}
\label{tab:statistics}
\end{table}

\begin{table}
\begin{center}
\begin{tabular}{|c|c|c|}
\hline
   Redshift   &Data set &  Power law fit \\
\hline 
 $5-3.5$  &Observed & ($-4.0$, $1.2$) \\
  $4$	 & DEF 	&($-5.3$, $2.1$) \\
 $3.5-2.5$	  &Observed & ($-4.1$, $1.5$) \\
  $3$	 & DEF 	&($-4.7$, $1.8$) \\
 $2.5-1.7$	  &Observed & ($-4.0$, $1.5$) \\
  $2$	 & DEF 	&($-4.0$, $1.7$) \\
\hline
\end{tabular}
\end{center} 
\caption{The coefficients of a power law fit to the velocity width and metallicity; ($\alpha$, $\beta$) from equation (\protect\ref{eq:powerlaw}).
}
\label{tab:powerlaw}
\end{table}

\begin{figure*}
\includegraphics[width=0.33\textwidth]{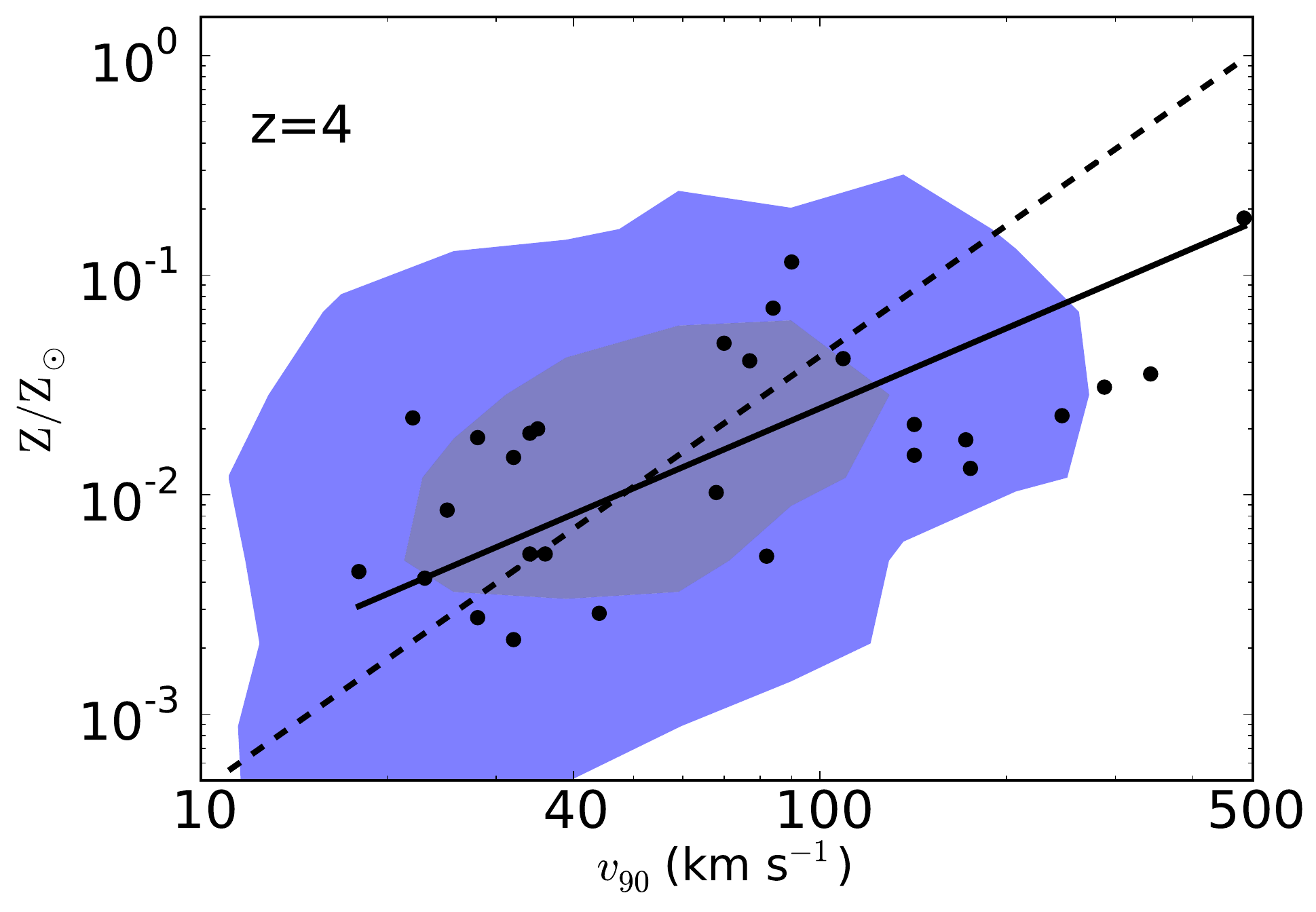}
\includegraphics[width=0.33\textwidth]{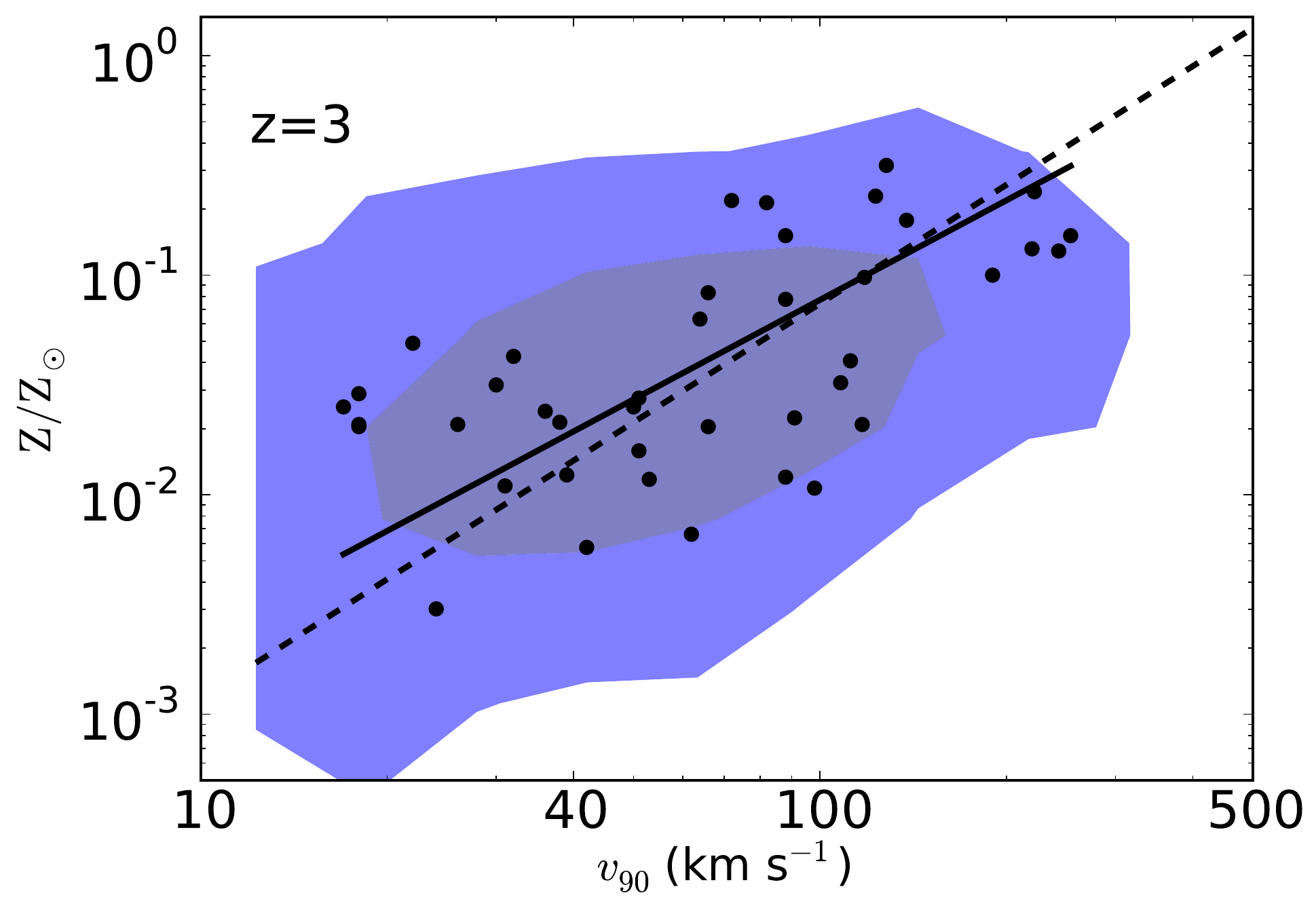}
\includegraphics[width=0.33\textwidth]{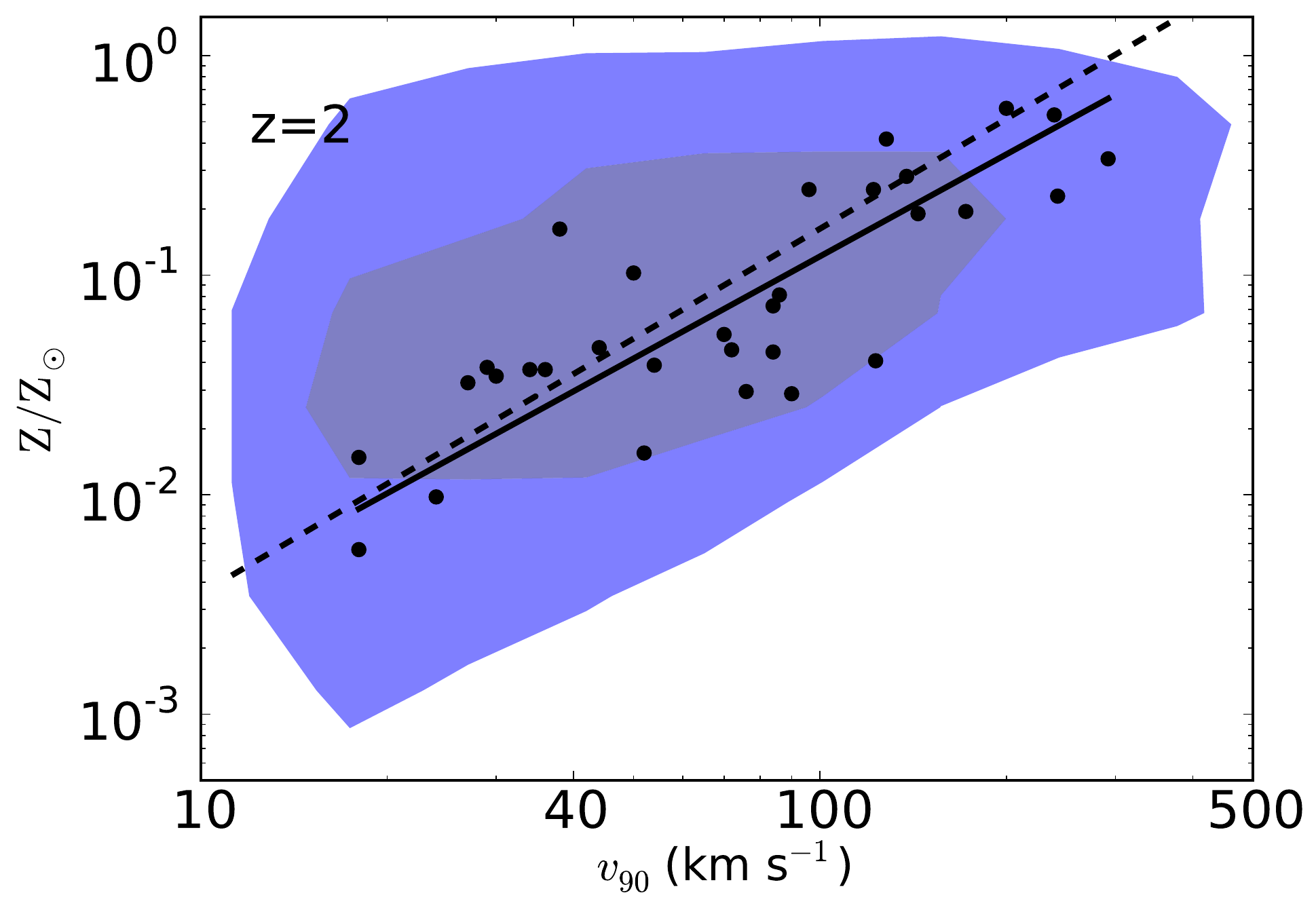}
\caption{Evolution of the velocity width metallicity plane from $z=4$ to $z=2$ for our DEF simulation, compared to observations
and with a linear fit superimposed. The ellipse shows a 2D histogram of the simulated data. 
The light blue region marks the area with $ > 0.15 \times 5000 / 40 $
sight-lines per grid cell, and the dark blue region the area with $ > 1 \times 5000 / 40 $. 
Solid lines show a power-law fit to the data points shown, and dashed lines a power law fit to the simulated spectra.
}
\label{fig:correlationzz}
\end{figure*}

Figure \ref{fig:correlationzz} shows the correlation from the DEF simulation, together with the observed 
data points and the power law fit. We showed in \cite{Bird:2014} that our preferred models reproduce the DLA metallicity very well.
The light blue region marks the area with $ > 0.15 \times N_\mathrm{sim} / N_\mathrm{obs}$
sight-lines per grid cell, and the dark blue region the area with $ > 1 \times N_\mathrm{sim} / N_\mathrm{obs}$.
$N_\mathrm{obs} \sim 40 $ is the number of observed sight-lines in each redshift bin, while $N_\mathrm{sim} = 5000$ 
is the number of simulated sight-lines. Thus the contours show an estimate of the region within which we expect to observe sight-lines.
As the metallicity evolves with redshift, we have split the data into three redshift bins, $z=3.5-5$, $z=2.5-3.5$ and $z=2.5-0$, and compared to simulated 
data at $z=4$, $3$ and $2$, respectively.

The simulated correlation is a good match to observations. Essentially all the observed data points lie within the 
larger of the shown contours. The only minor exception is three DLAs 
with an anomalously large velocity width in the highest redshift bin. It might be tempting to regard these objects as a signature of reionization; 
when substantial fractions of the Universe are neutral, DLAs will form in lower density, lower metallicity gas \citep{Rafelski:2014}. 
However, these DLAs are at $z \sim 4$, making this unlikely.
Instead, they are probably observational examples of spectra intersecting a low halo mass DLA and an LLS aligned along the line of sight, which
would produce a large velocity width with a below-trend metallicity. A by-eye examination reveals that three of these absorbers
contain two clearly separated peaks in the optical depth, as we would expect in this case.

\begin{figure*}
\includegraphics[width=0.33\textwidth]{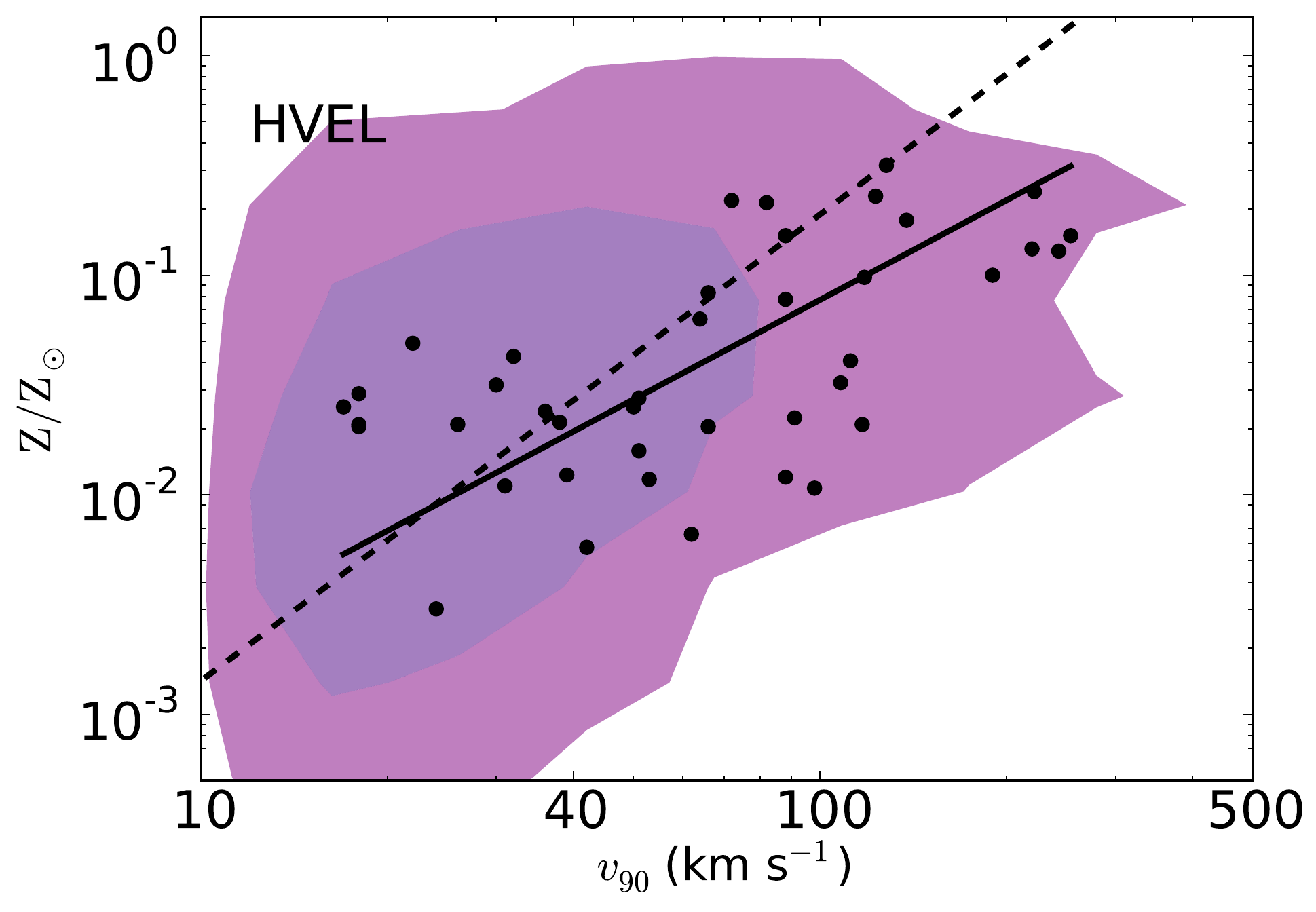}
\includegraphics[width=0.33\textwidth]{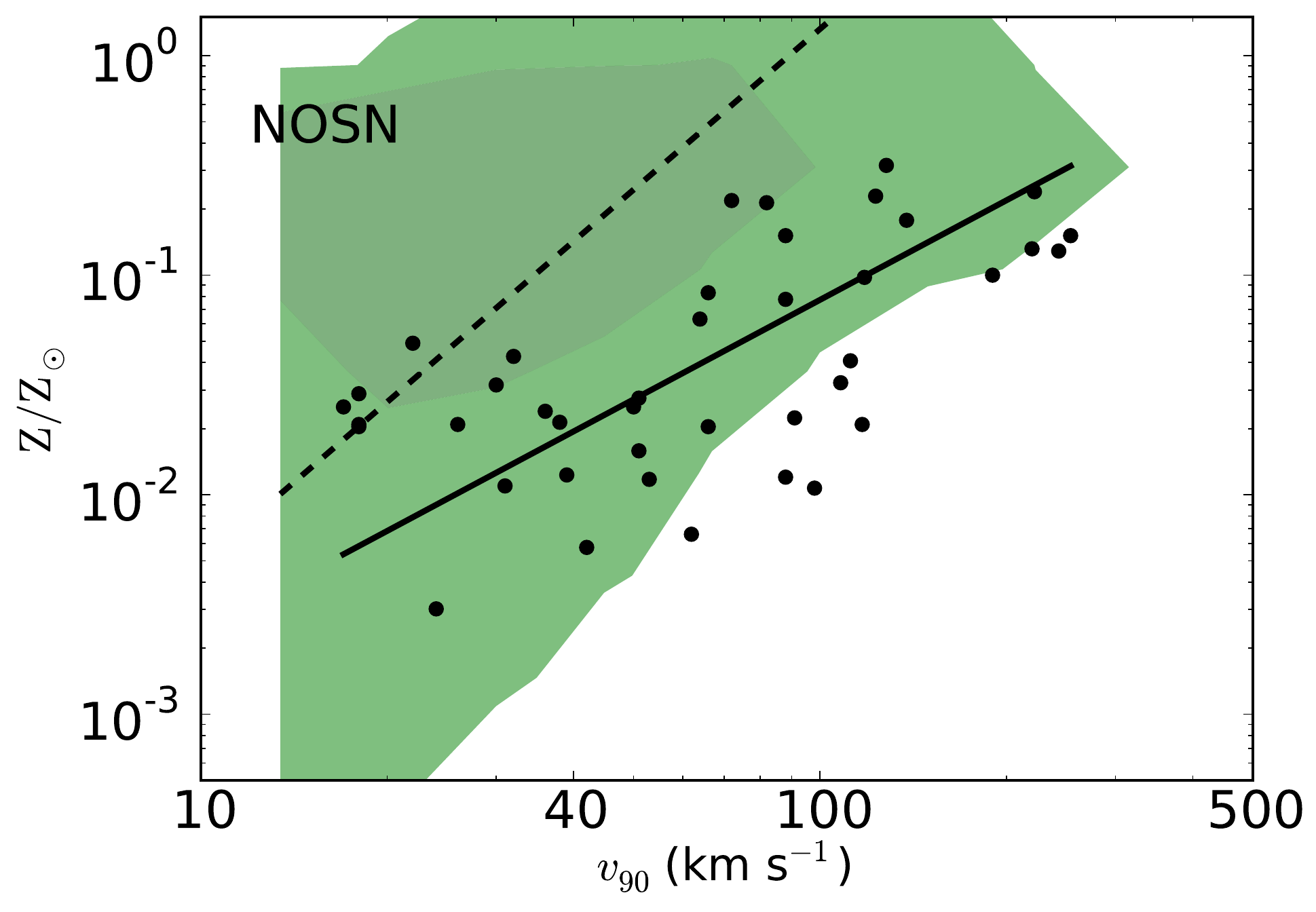}
\includegraphics[width=0.33\textwidth]{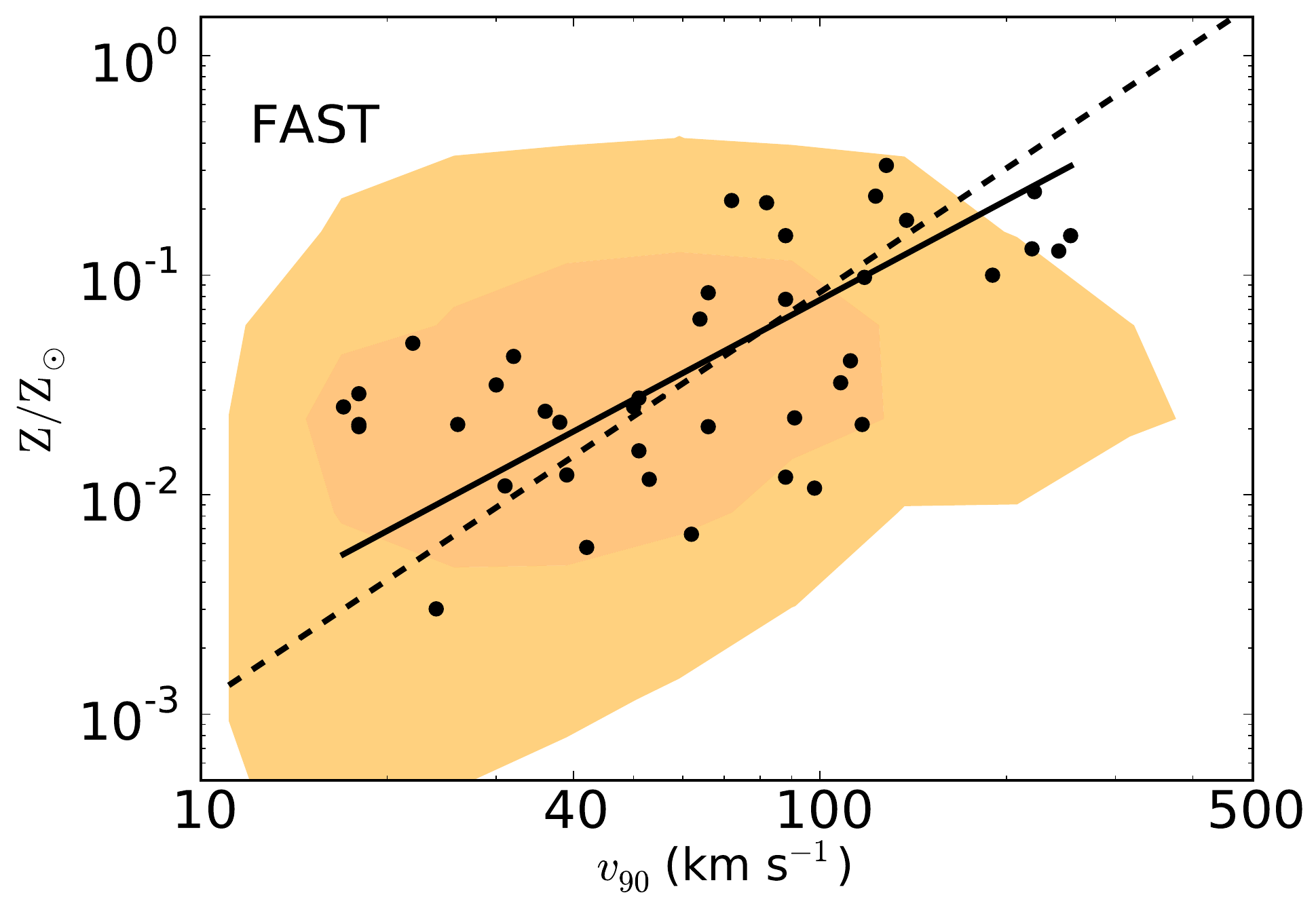}
\caption{Velocity width metallicity plane at $z=3$ for the three simulations not 
shown in Figure \protect\ref{fig:correlationzz}, with the best-fitting linear curve superimposed, and compared to observations.
Left: for HVEL, centre: for NOSN, right: for FAST. Light coloured regions marks the area with $ > 0.15 \times 5000 / 40 $
sight-lines per grid cell, and dark coloured regions the area with $ > 1 \times 5000 / 40 $.
Solid lines show a power-law fit to the data points shown, and dashed lines a power law fit to the simulated spectra.
}
\label{fig:correlations}
\end{figure*}

We have performed several statistical tests to determine whether the simulated and observed data sets are consistent, 
shown in Table \ref{tab:statistics}. We computed the Pearson $r$ coefficient between the metallicity and the velocity width, to measure the strength of the correlation.
For DEF we found $r = 0.4-0.5$, a little less than the observed value of $0.6$.
To determine whether the simulated and observed populations 
are drawn from the same population, we used the 2D Kolmogorov-Smirmov (KS) test \citep{Peacock:1983}. 
The KS test measures the maximal difference between the cumulative distribution function of two data sets; 
a larger value means they are less likely to be drawn from the same population.

As the correlation is relatively weak, the KS test may show poor agreement purely through statistical fluctuations.
To account for this, we re-calibrated the KS test using sub-samples of the simulations.
We randomly selected $50$ sub-samples of the simulated data, of the same size as the observed data set, and 
computed the 2D KS test score between this sub-sample and the full sample.
This allows us to compute the probability of the KS test value between the full sample and the observations being due to statistical fluctuations.
The 2D KS test between the DEF simulation and the data shows acceptable values at all redshifts. 
At $z = 3-4$, $30-40\%$ of mock trials produced a larger KS test, indicating that the observed 
data set is entirely consistent with the simulations. At $z=2$, agreement is more marginal. The high KS test value appears to be 
driven by sight-lines with velocity widths $\velwm \sim 20-40$ \kms~and metallicity above that expected from the linear correlation, 
suggesting that star formation may need to be further suppressed at $z < 3$, as also indicated by the HI abundance \citep{Bird:2014}.

We follow \cite{Pontzen:2008} in fitting a power law to the simulated and observed data set, 
using the least squares bisector described in \cite{Isobe:1990}. The power law has the form
\begin{equation}
 \frac{Z}{Z_\odot} = 10^\alpha \left(\frac{\velwm }{\mathrm{km\,s}^{-1}}\right)^\beta\,.
\label{eq:powerlaw}
\end{equation}
Note that the correlation is not strong, and the data set small. The fit does not capture the full behaviour of the data, 
and is sensitive to small changes in the data set used. We include it mainly as an aid to comparing the data sets, 
and recommend against using it in semi-analytic modelling. Our results are shown in Table \ref{tab:powerlaw}; DEF was in good 
agreement with the observations except at $z=4$, where the slope of the observed power law is reduced by the four anomalously low metallicity 
points discussed above.

Figure \ref{fig:correlations} shows the correlation at $z=3$ for our other simulations.
Both NOSN and HVEL are in clear disagreement with the observed data. HVEL does not match the observed 
velocity width distribution, and NOSN matches neither the observed velocity width distribution nor the observed metallicity.
Furthermore, in both cases the slope of the metallicity - velocity width relation is too steep. Unsurprisingly, the KS test scores 
for both simulations are very high, indicating agreement with the null hypothesis that the data sets are discrepant. Finally, the correlation between metallicity and velocity 
width is weaker in both these simulations. FAST is in marginal agreement with observations.
The velocity width distribution has too many low velocity width systems, producing a low KS test score, but
the correlation between metallicity and velocity width is still roughly reproduced.

\section{Conclusions}
\label{sec:conclusions}

We have examined the kinematics of Damped \Lya systems (DLAs) using cosmological hydrodynamic simulations. 
Our simulated DLA population resides in haloes with a mean virial velocity of $\sim 70$ \kms. 
We match the distribution of velocity widths of associated low ionization metal absorbers substantially better than previous work,
although on average our simulated velocity widths are still $20\%$ smaller than observed, implying 
that observed DLAs may be hosted in slightly more massive haloes than in our simulations.

Our much improved success in matching the velocity width distribution is mainly due to two features of our simulations.
First, our ionization modelling allows \SiII~to persist at relatively low densities, allowing
a larger fraction of the halo to contribute significantly to the absorption. 
Second, our supernova feedback models include strong outflows which suppress the formation of DLAs, especially in smaller haloes.
The largest velocity widths in our simulation exceed the virial velocity of any of the haloes, and are instead 
produced as a result of sight-lines which intersect a DLA and a LLS along the line of sight. 
To produce these systems, LLSs are enriched at the level of $\sim 1/100$ solar.

Our preferred feedback model reproduces the observed correlation between velocity width and metallicity, 
demonstrating that our simulations produce a realistic integrated star formation rate over a wide range of halo virial velocity.
There is some indication for a discrepancy at $z=2$, which is consistent with an excess of HI found in \cite{Bird:2014}.
We reproduce the observed equivalent width distribution of \SiII~$1526$ \AA, although this is not a strong 
constraint on our models.

The simulations in best agreement with the velocity width distribution exhibit a greater fraction of very edge-leading spectra, where 
the strongest absorption is at the edge of the profile, than observed. This discrepancy may be a statistical fluctuation in the data, but, 
if real, we speculate that it is due to the large mass of outflowing material produced by supernova feedback in our simulations.
While some outflows are necessary to enrich the circumgalactic gas, a model which augments them with another 
mechanism for suppressing star formation, such as thermal heating, might be better able to match both statistics.

We have searched for signatures of rotating discs in the simulated DLA population, looking at the relationship between virial 
velocity and velocity width as well as directly at the velocity structure of the cells causing the absorption. No more than $10\%$ of cells appear 
to be rotationally supported, and the relationship between virial velocity and velocity width predicted by rotating discs is at variance 
with that in our simulation. We thus conclude that at $z=3$ the fraction of rotationally supported gas in DLAs in our simulations is relatively minor.

Our improved match of the velocity width distribution uses simulations with the same stellar feedback implementation
which \cite{Bird:2014} showed was preferred in a comparison to the incidence rate and metallicity distribution of DLAs, 
thus further corroborating the conclusion of \cite{Bird:2014} that the DLA population at $z=3$ is hosted in haloes with moderate masses 
and typical virial velocities  of $\sim 70$ \kms.

\appendix

\section{Rotating Disc Models}
\label{ap:disc}

Following \cite{Prochaska:1997}, consider a thin rotating disc containing gas with a purely azimuthal velocity given by
\begin{equation}
 v(r) = v_\mathrm{vir} \frac{r}{R_\mathrm{vir}}\,,
\end{equation}
and a density profile
\begin{equation}
 n(r, z) \sim \exp\left(- \frac{r}{R_D} - \frac{|z|}{h}\right)\,.
\label{eq:discdens}
\end{equation}
The disc is entirely composed of baryons, while the mass of the halo is provided by dark matter. 
To set the scale height, assume conservatively that $90\%$ of the baryons are within 
the virial radius of the halo, which gives
\begin{equation}
 R_\mathrm{vir} = R_D \log 10\,.
\label{eq:discrad}
\end{equation}
Now consider a sight-line which intersects the mid-plane of the disc with impact parameter $b$, 
choosing coordinates so the x-axis goes through the intersection point.
We define $\phi$ to be the angle between the outgoing radial vector and the sight-line. The inclination angle $\iota$
is the angle between the sight-line and the plane of the disc. 
The velocity parallel to the sight-line is $v_\mathrm{vir} \sin (\phi) (r / R_\mathrm{vir}) $. At a height $z$ from the disc, the radius is
\begin{equation}
 r = \sqrt{b^2 + \frac{z^2}{\tan^2 \iota} + \frac{2 b z \cos \phi}{\tan \iota}}\,.
\end{equation}
For a thin disc we will have $z \ll b$
everywhere except a small area near the disc centre\footnote{If $z \gg b$, the analysis proceeds similarly, but with $\phi = 0$.}, and thus
\begin{equation}
 r \approx b + z \frac{\cos \phi }{\tan \iota}\,.
\end{equation}
The expected velocity width is the change in velocity between a height $H$ and $-H$ during which the sight-line intersects $90\%$ 
of the total material it will encounter. Integrating equation (\ref{eq:discdens}) along the sight-line we find that
\begin{equation}
 H = \frac{h R_D \tan \iota }{ R_D \tan \iota + h \cos\phi} \log 10\,.
\end{equation}
Using equation (\ref{eq:discrad}), the change in velocity will be 
\begin{align}
 \velwm &=  v_\mathrm{vir} \frac{\delta r}{R_\mathrm{vir}}  \\
	&= \frac{2 v_\mathrm{vir} \sin \phi}{1 + (R_D \tan \iota) / (h \cos \phi )}\,.
\label{eq:discvvir}
\end{align}
For a random sight-line, $\iota$ is uniformly distributed between $0$ and $\pi/2$ and $\phi$ uniformly distributed between $0$ and $2 \pi$.
The ratio of the disc scale length to scale height, $r_l = h/ R_D$, is a free parameter. Larger values will produce a distribution peaking at 
larger velocity widths. Figure \ref{fig:v90halomass} shows the resulting ratio between velocity width and virial velocity, generated 
by Monte Carlo sampling equation (\ref{eq:discvvir}) with $r_l = 0.25$, 
which is the largest value consistent with a thin disc approximation.

\section{Comparison to Previous Analyses}
\label{ap:compare}

\begin{figure}
\includegraphics[width=0.45\textwidth]{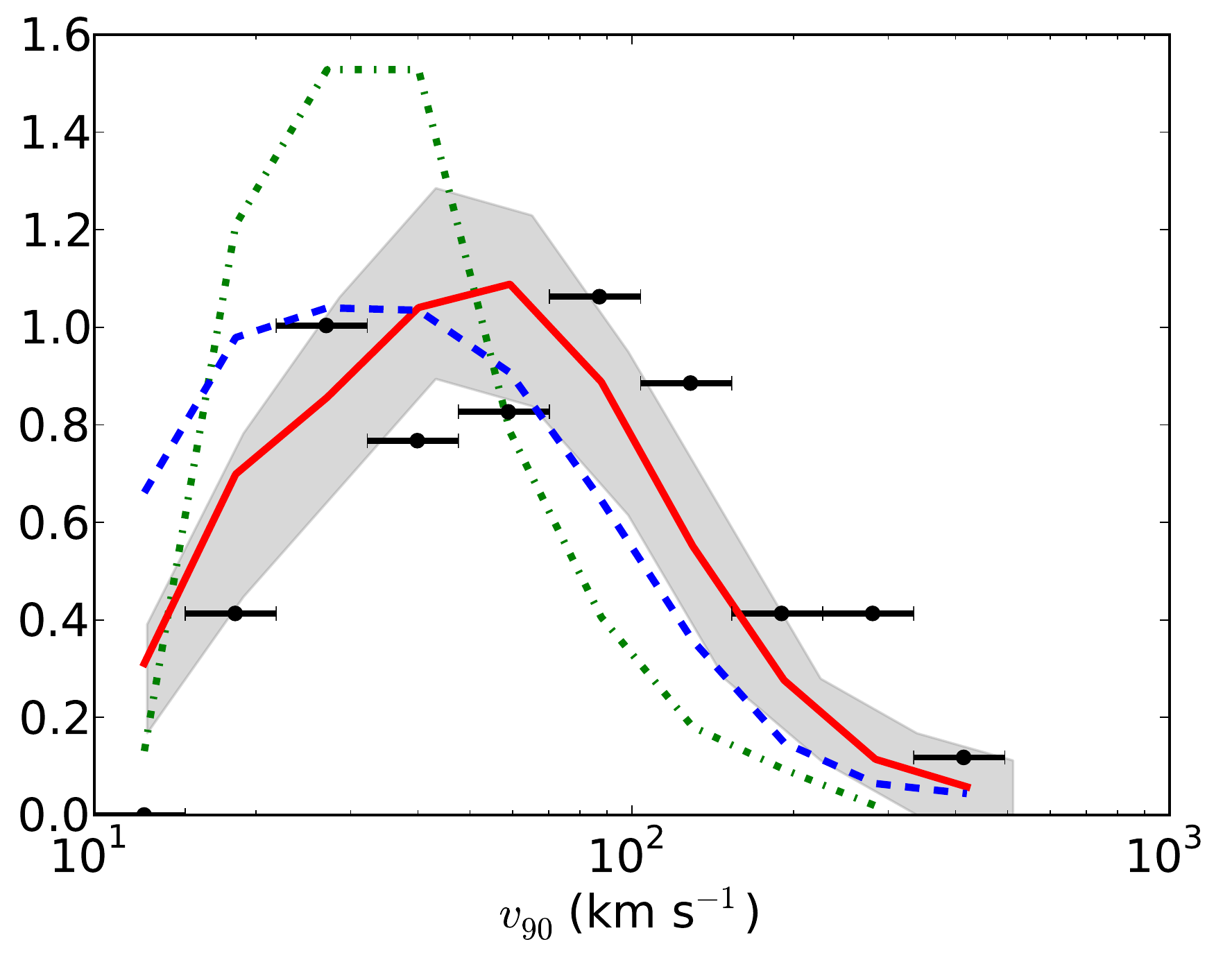}
\caption{The velocity width distribution at $z=3$ from the DEF simulation. Solid red shows 
our {\small CLOUDY} based analysis, dashed blue shows the results assuming that 
$n(\mathrm{SiII})/n(\mathrm{Si}) = n(\mathrm{HI})/ n(\mathrm{H})$, following \protect\cite{Pontzen:2008}. 
Finally, dot-dashed green shows an analysis following \protect\cite{Tescari:2009}.}
\label{fig:SiHI}
\end{figure}

In this appendix, we will examine why some recent analyses of DLA kinematics were not able 
to reproduce the high velocity width DLA systems. We shall focus on the analyses of \cite{Tescari:2009} and \cite{Pontzen:2008}, 
as they both used simulations which attempted to produce a cosmologically representative DLA population, rather than 
individual haloes. Comparing first to \cite{Tescari:2009}, two aspects of their analysis 
differ substantially from ours; their treatment of self-shielding and their selection of sight-lines.
Observed quasar sight-lines intersect foreground structure at random positions, whereas in their analysis sight-lines run through the 
centre of the $1000$ most massive haloes in their simulation, producing a bias towards larger velocity widths.
Self-shielding in their analysis occurs at the star formation threshold, $0.1$ cm$^{-3}$, an order of 
magnitude larger than the value we use. The \SiII~region of each halo is thus significantly smaller, 
biasing their velocity width distribution to low velocity widths. We re-analysed our simulation including both these effects, 
and found that the self-shielding threshold dominates. Using this analysis, the velocity width distribution peaks at $30$ \kms, 
and exhibits a lack of high velocity width systems, as shown in Figure \ref{fig:SiHI}.

\cite{Tescari:2009} computed the optical depth using a different method to ours.
They divided the sight-line into velocity space pixels, and each particle was interpolated on to the sight-line using a smoothing kernel, 
thus computing for each pixel an average density, temperature and velocity.
Each pixel is considered to be an absorber whose state is described by these averaged quantities, and the total optical depth computed 
using the Voigt profile. While adequate for the low densities and adiabatic evolution of Lyman-$\alpha$ forest absorbers, this
method loses small-scale velocity information, and is thus inaccurate when applied to narrow metal lines in high density systems.
To see why, consider two close particles moving with equal and opposite velocities;
averaging their properties to a velocity space bin will produce absorption in a single bin with zero velocity broadening,
whereas interpolating each cell directly as we do will produce the more realistic answer of absorption spread across bins with a velocity broadening
corresponding to the velocity of the original cells. This is probably the reason for the unphysical population
of very narrow lines with a width $< 2 $ km/s seen by \cite{Tescari:2009}.

\cite{Pontzen:2008} assume that the \SiII~abundance traces the neutral fraction of hydrogen, and 
so $n(\mathrm{SiII})/n(\mathrm{Si}) = n(\mathrm{HI})/ n(\mathrm{H})$. Figure \ref{fig:SiHI} shows the results of analysing our simulations using 
this assumption. Since it neglects the difference between the \SiII~and HI ionization potentials, 
it produces somewhat more compact \SiII~regions, systematically reducing the velocity widths. 
\cite{Tescari:2009} also computed the velocity width distribution using 
this formula for the \SiII~abundance, and found similar results.
One discrepancy remains; the simulation of \cite{Pontzen:2008} did not yield sight-lines with velocity widths larger than $250$ \kms.
This is because their analysis traced sight-lines only through the outskirts of 
haloes, and so did not produce sight-lines intersecting two widely separated absorbers, which we find make up a large fraction of these objects.

One final possibility is that our simulations are run using a moving mesh technique, 
which includes self-consistent advection of metals between cells. This will induce a smoother gas distribution 
and may also have improved the agreement of our results with observations. However, the large differences in analysis techniques
present in the literature mean that we are unable to draw strong conclusions about the magnitude of this effect 
without running our own smoothed particle hydrodynamics simulations with similar feedback and chemical enrichment models, which is beyond the 
scope of this work.

\section*{Acknowledgements}

SB thanks Ryan Cooke, Edoardo Tescari, Andrew Pontzen, Rob Simcoe and J.~Xavier Prochaska 
for useful discussions, and Volker Springel for writing and allowing us to use the \arepo~code. 
SB is supported by the National Science Foundation grant number AST-0907969, 
and the W.M. Keck Foundation. MN is supported by NSF grant AST-1109447.
MGH acknowledges support from the FP7 ERC Advanced Grant Emergence-320596. 
LH is supported by NASA ATP Award NNX12AC67G and NSF grant AST-1312095.

\label{lastpage}
\bibliography{DLAfeedback}
\end{document}